# Exoplanet Geology: What can we learn from current and future observations?

Bradford J. Foley
*Department of Geosciences, Center for Exoplanets and Habitable Worlds*
*Pennsylvania State University*
*University Park, PA 16802, U.S.A.*
*bjf5382@psu.edu*

## OVERVIEW

Nearly 30 years after the discovery of the first exoplanet around a main sequence star, thousands of planets have now been confirmed. These discoveries have completely revolutionized our understanding of planetary systems, revealing types of planets that do not exist in our solar system but are common in extrasolar systems, and a wide range of system architectures. Our solar system is clearly not the default for planetary systems. The community is now moving beyond basic characterization of exoplanets (mass, radius, and orbits) towards a deeper characterization of their atmospheres and even surfaces. With improved observational capabilities there is potential to now probe the geology of rocky exoplanets; this raises the possibility of an analogous revolution in our understanding of rocky planet evolution. However, characterizing the geology or geological processes occurring on rocky exoplanets is a major challenge, even with next generation telescopes. This chapter reviews what we may be able to accomplish with these efforts in the near-term and long-term. In the near-term, the James Webb Space Telescope (JWST) is revealing which rocky planets lose versus retain their atmospheres. This chapter discusses the implications of such discoveries, including how even planets with no or minimal atmospheres can still provide constraints on surface geology and long-term geological evolution. Longer-term possibilities are then reviewed, including whether the hypothesis of climate stabilization by the carbonate-silicate cycle can be tested by next generation telescopes. New modeling strategies sweeping through ranges of possibly evolutionary scenarios will be needed to use the current and future observations to constrain rocky exoplanet geology and evolution.

## INTRODUCTION

Exoplanets have revolutionized our understanding of the formation of planetary systems, continuing a long progression of scientific thought from ancient ideas of geocentrism towards the modern understanding of Earth's place in the galaxy and universe (e.g., see Dick 1993, for a historical review). In the early twentieth century it was unclear whether planetary systems were common or rare, formed only by fluke events like stellar close encounters as Jeans (1919) argued. By the mid-twentieth century, the sheer number of stars and prominence of the nebula collapse theory for planet formation popularized the view that planetary systems should be relatively common (e.g. Dick 1993). Before the discovery of the first exoplanets, though, we lacked any broader context for what planetary systems could look like, so our solar system was often assumed to be the default (e.g. Pfalzner et al. 2015; Dawson & Johnson 2018). However, with > 5000 confirmed exoplanets now discovered, it has become clear that our solar system is not the default for planetary systems. The first exoplanets discovered around sun-like stars were a type not seen in our solar system and one that strongly challenged models of planet formation: "hot Jupiters," gas giant planets orbiting very close to their host stars (Mayor & Queloz 1995). The radial velocity method, used to detect the first exoplanets and still one of the most fruitful methods used today, is biased towards finding large planets that orbit close to their host star, so close-in gas giant planets, if they existed, were going to be the first planets discovered. Work on exoplanet population statistics has now shown that hot Jupiters are not common, with frequencies of $\sim$ 1 % (e.g. Wright et al. 2012; Dawson & Johnson 2018; Zhu & Dong 2021). However, that they exist at all demonstrates that gas giant planets are not confined to the outer regions of planetary systems, as in our own.

In addition to hot Jupiters, there are other types of planets commonly found in exoplanet systems that are absent in our solar system. "Super-Earths," predominantly rocky planets that are larger in mass and radius than the Earth, and "mini-Neptunes" (or "sub-Neptunes"), planets larger than Earth but smaller than the ice giants in our solar system, and with a bulk density much lower than expected for a rocky planet, are common (e.g. Batalha 2014; Zhu & Dong, 2021; Kane et al. 2021). The mini-Neptunes may be ice and volatile rich or possess thick $H_2/He_2$ atmospheres (e.g. Bean et al. 2021). Exoplanet demographics show that the super-Earths and mini-Neptunes make up separate and distinct planet populations, separated by a radius "gap" at around 1.5-2.0 Earth radii (Fulton et al. 2017). That is, planets with a radius of $\approx$ 1.5-2.0 Earth radii are rare; planets are either smaller and rocky or larger and volatile rich and/or possess thick atmospheres comprised of nebular gas (e.g., see Bean et al. 2021, for a recent review). Exoplanets also show a wide range of system architectures compared to that seen in our solar system, further highlighting that ours is not the default structure for planetary systems (Kane et al. 2021; Zhu & Dong 2021).

Now armed with a broad statistical sample of the diversity of planets and planetary systems, the theory of planet formation is undergoing significant revision and new modeling approaches are being enabled (e.g. Drazkowska et al. 2023). As planet formation is an inherently stochastic process, studies of the formation of our solar system haven typically taken a statistical approach, running a large suite of models each with slightly different initial conditions, and looking at the statistical distribution of planets formed (e.g. Chambers 2001). Much has been learned from this approach on the proto-planetary disk characteristics and processes that can result in systems like ours. Naturally, though, we can be stuck with "just so" stories when trying to explain only our solar system. In fact, given the stochastic nature of planet formation, at least some aspects of our solar system inevitably are the result of specific events that occurred during formation, and therefore really are explained by seemingly ad hoc hypotheses. The small size of Mars may be an example of this. The current leading model is the "Grand Tack," where inward migration of Jupiter scattered solids away from the region where Mars would later form, leaving less mass available and hence a small planet (Walsh et al. 2011). Whether this (if it indeed is the correct explanation for Mars's size) is a general process that would happen broadly in planetary systems, or just a result of the random quirks that went into forming our own system is not currently known. However, exoplanets provide a whole population of planets and systems for models to try to match. With this population it will be easier to work out which aspects of planetary systems are a result of general physics and processes occurring during formation, and which are due to random chance. Population synthesis studies are embarking on this line of research now and are already reshaping our understanding of planet formation (e.g. Mordasini et al. 2009).

The work of discovering exoplanets and quantifying the demographics of planetary systems will of course continue for years to come and is vital to answering the question of whether our solar system architecture is rare or common. However, with the launch of the James Webb Space

Telescope (JWST), and plans for future missions, such as the direct imaging Habitable Worlds Observatory (HWO) or the proposed mid-infrared Large Interferometer For Exoplanets (LIFE) (Quanz et al. 2022), we are now entering an era of better characterization of exoplanets. With these new observational capabilities comes new opportunities to constrain not just basic planet properties like size and bulk density, but to probe their atmospheres or even surfaces. These new observations can then potentially be used to infer something about the geological characteristics, such as the surface lithology, or geological processes, such as tectonics or volcanism, of rocky exoplanets.

As a result, exoplanets have the potential, at least seemingly, to spur a revolution in our understanding of planetary geological evolution, the same way they have revolutionized our understanding of planetary system formation and architecture. Like the field of planet formation before the discovery of exoplanets, geoscientists and planetary scientists only have a limited sample of planets and moons to study in detail: there are only four rocky planets (the topic of this volume) in our solar system. Each of these planets has been studied well enough to constrain to first order surface and interior composition, surface features, and tectonic processes operating today, though of course many details remain topics of intense study and debate. The four rocky planets of our solar system show significant diversity: a range of sizes from Mercury to Earth, different core sizes, tectonic states, atmosphere sizes and compositions, and magnetic fields or lack thereof for Venus and Mars. Earth is also the only planet where plate tectonics is known to operate. Mars and Mercury are likely in a "stagnant lid" mode of tectonics, where convection operates in the mantle but is unable to "break" the lithosphere into discrete plates that can move with respect to each other (e.g. Breuer & Moore 2015; Bercovici et al. 2015). Venus does not possess a global network of mobile plates like the modern Earth does, as evidenced by the lack of hallmark topographic features like ridges at zones of plate divergence or subduction zone trenches at regions of convergence. Venus may not be in a stagnant-lid state, however, as it does show evidence for localized subduction (Sandwell & Schubert 1992; Davaille et al. 2017) and regions of thin lithosphere and high heat flow that are not consistent with stagnant lid convection (Borrelli et al. 2021; Smrekar et al. 2023). Venus may therefore operate in a tectonic regime intermediary to plate tectonics and stagnant-lid tectonics.

Geoscientists and planetary scientists have sought to explain the differences between the current states and evolutionary histories of the four rocky planets for decades. In particular, how Earth evolved into a habitable planet with liquid water, a temperate climate, plate tectonics, and a magnetic field, while the other rocky planets in our solar system did not, has been a long-standing question. Further motivating this question is the potential importance of both plate tectonics (e.g., Kasting & Catling 2003) and the magnetic field (e.g., Cockell et al. 2016) for Earth's habitability. Much progress has been made on elucidating the basic mechanics behind key processes and characteristics of the solar system rocky planets, like the physics behind plate tectonics and mantle convection, generating a magnetic field, or atmospheric evolution and retention.

However, as with solar system formation, explanations of Earth's fundamentally different evolution in comparison to its neighboring planets can ultimately lead to similar "just so" stories, due to the lack of a broader sample of planets with which to compare the Earth. With only one planet as an example of a habitable planet with plate tectonics and a magnetic field, and only a handful of counter examples, it is hard to generalize and test theories for why Earth ended up the way it did. Multiple hypotheses can explain the scant available data, and it is hard to disentangle cause and effect for some of Earth's unique features. For example, is plate tectonics caused by the presence of liquid water oceans, through the rheological weakening effects of water (Tozer 1985; Mian & Tozer 1990; Lenardic & Kaula 1994; Moresi & Solomatov 1998; Richards et al. 2001; Regenauer-Lieb et al. 2001; Korenaga 2007, 2010), or are liquid water oceans themselves caused by plate tectonics, and the climate stabilization it helps to provide through the carbonate-silicate cycle (Landuyt & Bercovici 2009)? A larger sample of planets, if they can be well characterized, could help answer these questions.

However, characterizing geological processes or properties of exoplanets is a significant challenge. Therefore, the ability of exoplanet studies to revolutionize our understanding of rocky planet diversity and evolution is far less certain than for planetary system formation. Finding exoplanets and constraining their sizes and orbits was enough to completely change our view of how planetary systems formed. For geological processes, far more detailed information about exoplanets is needed. This chapter therefore reviews the prospects for constraining geological properties and processes on exoplanets, and how these constraints might affect broader understanding of rocky planet evolution. The chapter focuses on a few key areas and discusses current or future observations that might help constrain them. First the chapter examines how compositional diversity of exoplanets might be constrained using current or near future observations. Then it discusses how current and future observations of rocky exoplanet atmospheres can inform their geological processes. In particular, this section looks at current discoveries of planets that appear to lack atmospheres and reviews what the lack of an atmosphere implies about the release of volatiles from the interior ("outgassing") on such planets. The following section then looks at what we might learn from future direct imaging missions, focusing on how scientists might test whether the carbonate-silicate cycle operates on rocky exoplanets to regulate their climates, and what implications this has for rocky planet interiors. Finally, the chapter concludes with summary thoughts and discussion.

# TESTING MODELS OF THE COMPOSITIONAL DIVERSITY OF ROCKY EXOPLANETS

Rocky exoplanets can come in a wide variety of compositions, which will influence factors like the size of the core relative to the silicate mantle, the mineralogical makeup of the mantle and crust, as well as important material properties that control the dynamics of the mantle and core, how they evolve over time, and whether processes like plate tectonics or core dynamos can operate. However, directly constraining the composition of rocky exoplanets is clearly difficult, as we cannot directly sample and analyze their surface and mantle rocks. Several approaches have been developed to provide more indirect constraints on the plausible range of exoplanet compositions, and hence to supply geodynamicists with a framework to work within exploring how planet composition influences geophysical processes (e.g. Spaargaren et al. 2020). One approach is to construct interior structure models of planets with different compositions, and then use the measured mass and radius of individual planets to assess which compositional model they best match (Fortney et al. 2007; Valencia et al. 2007a,b; Seager et al. 2007; Zeng & Sasselov 2013). However, rocky planet composition cannot be uniquely inferred from mass and radius information alone (Dorn et al. 2015), so additional constraints are necessary. One popular avenue is to use star compositions to infer planet compositions (Dorn et al. 2015; Unterborn et al. 2016). As planets and the stars they orbit originally form from the same proto-stellar nebula, then planets should roughly match the composition of their host stars for refractory (rock-forming) elements (Thiabaud et al. 2015; Lodders 2020; Jorge et al. 2022). Measuring the composition of refractory elements in stars therefore provides an indication of the range of plausible rocky planet compositions. A number of studies have used compilations of stellar compositions, from databases such as the Hypatia catalog (Hinkel et al. 2014), to estimate this range of planet compositions, and construct models of resulting interior structures, mantle mineralogies, and heat budgets (Unterborn et al. 2015; Hinkel & Unterborn 2018; Putirka & Rarick 2019; Unterborn et al. 2022; Spaargaren et al. 2023). This approach of using stellar composition to estimate planet composition is covered in Xu et al. (20XX, this volume).

Using stellar compositions to infer planet composition of course relies on the assumption that planets will approximately match the composition of their host stars, which ultimately needs to be tested observationally. Determinations of resulting interior structure and mantle mineralogy also rely on models which inevitably have limitations and may not always capture reality. As such it is critical to find additional ways to constrain the composition of rocky exoplanets more directly. This section focuses on other observations that provide such a constraint. These observations can only be attained in certain situations, meaning they cannot be used to constrain the full range of compositions across the observed exoplanet population, the way mineralogical models based on stellar composition can. However, even if just a few planets can have their crust, mantle, or even whole planet bulk composition estimated from direct observations, this will serve as a valuable test of models based on stellar compositions.

**Direct measures of rocky planet composition**

Four potential avenues for more directly constraining the composition of rocky exoplanets are: 1) pollution of white dwarf stars; 2) measuring the chemical composition of dust grains derived from "disintegrating planets,"; 3) measuring the chemistry of silicate vapor atmospheres for planets with magma oceans at their sub-stellar points ("lava worlds") or the chemistry of the lava itself; and 4) estimating surface compositions from thermal emission of airless rocky bodies. White dwarf pollution is a powerful tool that provides a direct measure of the elemental abundances of rocky exoplanetary material falling onto ("polluting") a white dwarf star. This material is likely the remnants of rocky planets that once orbited the star. White dwarf pollution has already been used to constrain the redox state of exoplanetary silicate materials, finding states similar to Earth and Mars (Doyle et al. 2019); this result has important implications for exoplanet mineralogy, interior structure, and atmospheres. Veras et al. (20XX, this volume); Xu et al. (20XX, this volume) thoroughly discuss white dwarf pollution and what constraints observations so far place on exoplanet geology, so this section will focus on the latter three methods.

**Disintegrating planets**

Ultra-short period (USP) planets are planets that orbit their host star with extremely short orbital periods and hence extremely small orbital separations (Sanchis-Ojeda et al. 2014). Some rocky USP planets orbit so close to their host star, and therefore receive such high levels of stellar flux, that the rocky surface is evaporating and being lost to space (e.g. Rappaport et al. 2012; Sanchis-Ojeda et al. 2015; van Lieshout & Rappaport 2018). The result is a rocky planet with dust clouds that can either lead or trail the planet, or both, causing a distinct shape to the transit light curve that can also vary significantly over time (e.g. Rappaport et al. 2012; van Werkhoven et al. 2014; Sanchis-Ojeda et al. 2015). For vaporized silicate to escape to space, the planet itself must be small, on the order of Mercury-sized to the size of Earth's moon (Perez-Becker & Chiang 2013), and these are the sizes inferred for the 3 disintegrating planets found so far (Rappaport et al. 2012, 2014; Sanchis-Ojeda et al. 2015).

Disintegrating planets provide an excellent opportunity to constrain their composition, because the widely scattered dust grains around the planet are well suited for study with "transmission spectroscopy" (Bodman et al. 2018). In transmission spectroscopy, light from the star passes through a planet's atmosphere, or in this case dust clouds, during transit. Comparing the signal during transit and outside of transit allows the contribution from the planet to be isolated. In the case of disintegrating planets, the spectroscopic signal from the dust grains can be used to infer their composition (Okuya et al. 2020), with K2-22b probably the planet best suited for this analysis (Bodman et al. 2018). Interpreting the observed dust composition to constrain the composition of the disintegrating planet will not be without its challenges. It is not immediately clear if the crust, mantle, or core of the planet is disintegrating, or if dust will reflect some mixture of all three potential compositional layers. While the composition of the dust itself once measured can help shed light on this, as more silica-rich minerals would be expected from the crust, and significant iron abundance would indicate disintegration from the core, there will be some degree of degeneracy between the bulk planet composition and the presumed layer or layers that dust is derived from.

Disintegrating planets are rare and may only spend $\sim 10-100$ Myrs during the phase of active disintegration (Perez-Becker & Chiang 2013). Such planets will therefore not form a large dataset of planet compositions with which to test compositional models like those presented in Putirka (20XX, this volume). Furthermore, the origin of disintegrating planets is not well understood. Whether they were born as rocky planets in the inner proto-planetary disk (either in-situ or at least relatively close to where they are currently found) or formed further out as mini-Neptunes or even gas giant planets that have migrated to their current ultra-short period orbits and lost their gas envelopes is not clear (e.g. Jackson et al. 2013; Lopez 2017; Winn et al. 2018). Observations of the dust being lost from these planets can potentially shed light on this formation history as well, which is important for interpreting what they mean for rocky planet exoplanet compositions more generally. If disintegrating planets began their lives as mini-Neptunes or gas giants, then their composition may not be as reflective of rocky exoplanets more broadly. However, they nonetheless can help complement observations from e.g. white dwarf pollution and, when viewed in the proper context of their potential formation mechanisms, help test models of exoplanet compositional diversity.

**Lava worlds**

A less extreme type of USP planet than disintegrating planets are those that receive enough stellar radiation flux to have dayside temperatures exceeding the melting temperature of the crust or underlying mantle, leading to large lava ponds or magma oceans on their surfaces. However, stellar flux is not so high, or the planet mass is large enough that silicate vapor is not being lost to space at substantial rates, as in the case of disintegrating planets. These planets are "lava worlds" (see review by Chao et al. 2021), and have been recognized dating back to the discovery of some of the first rocky exoplanets (Léger et al.,2009). Some of the most well-known lava planets are CoRoT-7b (Léger et al. 2011), 55 Cancri e (Winn et al. 2011; Demory et al. 2011), Kepler-10b (Batalha et al. 2011), and Kepler-78b (Howard et al. 2013; Sanchis-Ojeda et al. 2013).

Theoretical models predict that lava worlds could have atmospheres composed of vaporized silicates, in equilibrium with the lava ponds or oceans at the surface of the dayside of these planets (Schaefer & Fegley 2009; Miguel et al. 2011; Schaefer et al. 2012; Ito et al. 2015). These silicate vapor atmospheres are expected to be dominated by Na, $O_2$, Si, SiO, or K (Schaefer & Fegley 2009; Ito et al. 2015), depending on the composition of the surface experiencing melting and vaporization (Miguel et al. 2011; Schaefer et al. 2012), though volatiles like Na and K may be lost leaving SiO or $SiO_2$ as dominant observable constituents (Nguyen et al. 2020; Ito & Ikoma 2021).

If the composition of these silicate vapor atmospheres can be measured, then they provide a direct window into the composition of the interior (Zilinskas et al. 2022; Piette et al. 2023). As a result, studying the atmospheres and geodynamics of lava worlds has become a major focus of activity, with papers considering details of atmospheric structure and transport (Hammond & Pierrehumbert 2017; Nguyen et al. 2022), dynamics of the magma ocean (Kite et al. 2016), and convection in the whole mantle (Meier et al. 2023).

55 Cancri e is the most well studied lava world. Demory et al. (2016) mapped thermal emission from the planet by measuring the host star's light curve as the planet passed around behind the star. As a transiting planet orbits towards its "secondary eclipse," where it passes directly behind the host star, more and more of the dayside of the planet comes into view from our reference frame. As a result, the observed light includes both the star's light and light emitted by the planet. However, when the planet reaches secondary eclipse, the planet's contribution drops away, leaving only the star's light. Comparison of the star plus planet emission throughout an orbit to that of just the star during secondary eclipse allows for the contribution from the planet to be isolated. Demory et al. (2016) found a large day/night temperature contrast, indicating limited redistribution of heat from day to night side. However, the point of peak emission was offset from the substellar point, indicating flow of either atmosphere or lava redistributing heat on the day side. The offset in peak emission has been interpreted as being indicative of an atmosphere (Angelo & Hu 2017; Hammond & Pierrehumbert 2017), though stronger confirmation is still lacking. In fact, reanalysis of the data by Mercier et al. (2022) finds a negligible peak emission point offset, indicating minimal planetary heat redistribution; clearly additional observation and analysis is needed. Using transmission spectroscopy, an abundant low mean molecular weight atmosphere can be ruled out, but these observations can't distinguish between clouds, a high mean molecular weight atmosphere like a silicate vapor atmosphere, or no atmosphere at all (Deibert et al. 2021). The lack of any evidence for escaping Helium also indicates that 55 Cancri e lacks a low mean molecular weight atmosphere today, either because it has been lost already or never accreted one to begin with (Zhang et al. 2021).

Observations with JWST will hopefully confirm the nature of 55 Cancri e and other lava worlds atmospheres one way or another. As of yet lava world atmospheres are still awaiting detailed characterization, and thus their ability to constrain rocky planet composition is still mostly theoretical. Zieba et al. (2022) measured thermal emission of a different lava world, K2-141 b, and found emission consistent with a thin silicate vapor atmosphere. This might be our first tantalizing glimpse at the characterization of lava world atmospheres, that hopefully will be refined in the coming years. Another possibility for constraining lava world compositions is using the emission or reflection of light from their magma or crystal mush surfaces (Fortin et al. 2022), similar to what can be done for solid surfaces as discussed in the next section. This work is also in its infancy but could provide another promising avenue for rocky planet compositional characterization.

**Emission from solid surface planets**

Moving back to larger orbital distances are planets cool enough to have solid surfaces, but that still receive such high radiation fluxes from their host stars that they lack atmospheres. Thermal emission is the primary observation used to look for these "airless" or "bare rock" planets. Planets lacking atmospheres should lack heat redistribution from dayside to nightside, resulting in high temperatures on the dayside and strong day-night temperature contrasts. Measuring the full thermal phase curve of a planet can show this temperature contrast and hence whether there is significant heat redistribution, though an even simpler method is to measure dayside thermal emission, constrain the temperature, and compare with models that account for heat redistribution by different assumed atmospheres and an airless case (Koll et al. 2019). A handful of planets that are good candidates to be bare rocks have been identified so far based on thermal emission observations, as outlined in more detail below in *Observations of rocky planet atmospheres to date.*

Bare rock planets present a rare opportunity for characterizing surface geology, since the observed thermal emission comes from the surface rocks. Different types of crust (e.g. mafic crust like the basalt making up Earth's seafloor, or felsic crust like our continents, predominantly formed of granitic-type rock) produce different emission as a function of wavelength. Observing thermal emission from a bare rock planet in different wavelengths can therefore allow the surface rock composition to be constrained (Hu et al. 2012).

Thermal emission observations have been made for planets thought to be bare rocks: LHS 3844b (Kreidberg et al. 2019; Whittaker et al. 2022), GJ 1252b (Crossfield et al. 2022), Trappist-1b (Greene et al. 2023; Ih et al. 2023), and Trappist-1c (Zieba et al. 2023). It should be noted, though, that the evidence supporting the inferred lack of atmosphere for these planets is of varying robustness, e.g. the observations of Trappist-1c are permissive of some hypothetical thin atmospheres (Lincowski et al. 2023). More observations and better models will be needed to confirm which planets are truly airless. Results obtained so far assuming these planets are indeed bare rocks have found surface compositions broadly consistent with mafic or ultramafic crusts, similar to the basalt ocean crust on Earth today. Another possible crust composition is a highly oxidized, hematite rich crust which would be produced by photolysis of water, escape of H to space and oxidation of the surface by remaining O (Hu et al. 2012). The results are not precise, and a wide range of compositions are consistent with the observations for each planet.

However, for the most part felsic crusts have been found to not match the data. Mafic or ultramafic crusts would be expected to form via partial melting of a mantle with a similar composition to Earth's, so the crust compositions inferred so far appear consistent with an Earth-like composition (see Purika, 20XX this volume; Shorttle and Sosi, 20XX this volume). Observations of a felsic crust would be exciting, as felsic crust on Earth predominantly forms due to hydrous melting (e.g. Campbell & Taylor 1983). Felsic crust on an exoplanet could indicate that water was present at least at some point during the planet's history, even if water has since been lost. Liquid water is not expected on planets receiving the high stellar fluxes that the candidate bare rocks do, but even unlikely possibilities should be considered and tested given the lack of definitive constraints. Felsic crust could potentially also form from mantle melting and distillation on planets with non-Earth-like (probably more silica-rich) bulk compositions (Putirka & Rarick 2019; Brugman et al. 2021). Combining stellar composition constraints and thermal emission where possible could therefore act as a valuable test of exoplanet composition models. In particular targeting planets where stellar compositions would imply a mantle that would produce something other than mafic or ultramafic crust could help confirm predictions from compositional models.

# CONSTRAINTS ON EXOPLANET GEOLOGY FROM THE PRESENCE OR ABSENCE OF AN ATMOSPHERE

One of the top priorities for exoplanet science, both currently and in the coming decades, is to characterize exoplanet atmospheres to constrain their prospects for habitability, look for potential signs of life, learn about planet formation and evolution, and even potentially constrain geological processes (e.g. National Academies of Sciences, Engineering, and Medicine 2021). Attempts to characterize rocky planet atmospheres are still in their infancy. The Hubble and Spitzer space telescopes have been used to look at a handful of rocky planets so far (see e.g. Wordsworth & Kreidberg 2022, for a review), and now with JWST we are entering a new era of rocky exoplanet atmospheric observing capabilities. Observations of Trappist-1 and other systems will hopefully provide our first close look at the atmospheres of habitable zone rocky planets.

Given the wealth of exoplanet atmosphere data that will be collected in the coming years, the science of exoplanet geology must be poised to utilize these observations to learn as much about rocky exoplanet interiors as possible to move forward. The challenge facing researchers interested in exoplanet interiors, however, is that atmospheres are only an indirect result of rocky planet interior processes, and thus only indirectly constrain these processes. Tectonic and geological processes on solar system planets are largely studied by detailed mapping of surface topography, composition, and geophysical quantities like gravity; such detailed surface characterization is not feasible in the foreseeable future for exoplanets. It will therefore be difficult to tightly or uniquely constrain geological processes based on atmospheric observations. However, given the wide diversity possible for exoplanets and the dearth of information about their geology, any constraint that can at least rule out some scenarios is still highly valuable.

In that spirit this section will focus on what can potentially be constrained about rocky exoplanet geological processes from the most basic, and robust, aspect of atmospheric characterization conducted so far, which is simply determining which planets do or do not have atmospheres at all. While the ultimate goal of rocky exoplanet atmospheric characterization is detailed mapping of the abundances of different atmospheric gasses, as outlined in more detail in *Observations of rocky planet atmospheres to date* below, this has proved challenging so far. However, a handful of rocky exoplanets planets have been inferred to various levels of confidence to be airless. These observations motivate the question of what factors control atmosphere formation and retention on rocky planets, a question which is likely to drive much of the work in the field in the near-term. A key aspect of atmosphere formation and retention is exoplanet geology, as volcanism and release of volatiles from the interior (called outgassing) is a major source of atmospheric gases for rocky planets. This section will therefore focus on the constraints provided on volatile outgassing by knowledge of whether a planet does or doesn't have an atmosphere.

Airless exoplanets also provide some information about the composition of rocks at the surface, integrated over the surface area from which emission on the planet is observed. This surface composition constraint provides a window into the overall planet composition (see *Emission from solid surface planets*) and also informs whether volcanism has taken place at least at some point in time during the planet's history as discussed more in this section. Future telescopes may also be able to constrain brightness variations across a planet's surface as it rotates through direct imaging, which can be used to estimate lateral compositional differences or the distinction between e.g. oceans and continents (e.g. Ford et al. 2001; Cowan et al. 2009; Zugger et al. 2010). Thus, surface mapping of exoplanets may be possible in the future. However, these very coarse constraints are the limit of exoplanet surface characterization for the coming decades, and yet are still far below the level of detail geologists and planetary scientists are used to when studying the geological processes on solar system planets. Hence the clear need to use atmosphere observations as at least a complementary tool to help constrain exoplanet geology.

This section will first briefly review the methods for atmosphere characterization, then observations that have been made so far, with a focus on the discovery of a number of presumably airless planets. It then reviews the key sources and sinks of atmospheric gases, volcanic outgassing, and atmospheric escape. The section closes by discussing how the observation of a lack of atmosphere on a planet can be combined with models of escape to constrain outgassing rates and evolutionary history.

**Methods for Atmosphere Characterization on Rocky Exoplanets**

Exoplanet atmospheres can be constrained by transmission spectroscopy and thermal emission, or in future missions by direct imaging (e.g. Wordsworth & Kreidberg 2022). Transmission spectroscopy and emission were briefly introduced in *Testing Models of the Compositional Diversity of Rocky Exoplanets* but will be more fully explained here. Transmission spectroscopy refers to measuring the starlight as it passes through a transiting planet's atmosphere, and comparing this to the starlight observed when the planet is not transiting (e.g. Seager & Sasselov 2000; Brown 2001; Ehrenreich et al. 2006; Kaltenegger & Traub 2009). From comparing these two cases, the signal from the planet's atmosphere can be isolated. In particular, measuring the flux in different wavelengths allows potential absorption features for key atmospheric gases to be detected, and hence characterization of the species making up the planet's atmosphere. Atmospheres can also be characterized by the thermal emission the plant gives off. Here the combined radiation emitted by both the planet and star during the planet's orbit is measured (e.g. Deming et al. 2005). For a transiting planet, the night side of the planet will be facing observers on Earth when it passes in front of the star, while the day side of the planet will be facing observers as it orbits behind the host star before the planet disappears behind the star during "secondary eclipse." Tracking how the combined star plus planet emission changes over the course of a planet's orbit allows the difference between day and night side temperatures to be constrained, which informs whether heat is being redistributed by an atmosphere or not (e.g. Knutson et al. 2007). Looking at emission in different wavelength bands further allows astronomers to compare observations with different atmosphere, or for airless planets, surface, composition models (e.g. Sudarsky et al. 2003; Hu et al. 2012; Stevenson et al. 2014; Kreidberg et al. 2019; Piette et al. 2022). Finally, next generation space telescopes will use direct imaging for atmosphere characterization, where light from the planet is directly observed. These telescopes will use choronographs to block out the star's light, allowing light from the planet to be observed and analyzed (National Academies of Sciences, Engineering, and Medicine 2021).

 Detailed characterization of rocky exoplanet atmospheres is challenging. Obtaining sufficient signal from the planet requires combining observations from multiple transits for either transmission spectroscopy or mapping thermal emission. Both techniques are also more successful for smaller, cooler stars as the planet will have a larger impact on the observed starlight in such cases. These practical considerations favor planets in short period orbits around M-dwarf stars as the most readily observable. Planets in such systems are more likely to transit from our reference frame, the short period orbits allow for multiple transits to be observed with limited observing time, and M-dwarf stars are abundant and frequently host rocky planets (e.g. Wordsworth & Kreidberg 2022). Moreover, the cool temperature of M-dwarfs means that the habitable zone is much closer to the star than for a sun-like star, so even habitable zone planets have short period orbits (e.g. Kasting et al. 1993b; Kopparapu et al. 2013).

 However, stellar contamination appears to be a major problem for transit spectroscopy for planets around M-dwarfs (e.g., Lim et al. 2023). Moreover, M-dwarf stars also differ from sun-like stars in ways that may be detrimental to habitability. M-dwarf stars take up to ~ 1 Gyr to reach the main sequence, and during this time they are highly luminous and emit high energy radiation. This high luminosity and extreme radiation can potentially desiccate the planets that will eventually end up in the habitable zone of these stars after they reach the main sequence (Ramirez et al. 2014; Luger & Barnes 2015). Planetary systems around M dwarfs after they reach the main sequence may therefore be left with only desiccated planets, including those in the habitable zone, and planets too cold to support liquid surface water beyond the habitable zone outer edge. Nevertheless, the favorable observing conditions for planets around M-dwarfs mean they have been the primary focus of attempts to characterize rocky exoplanet atmospheres thus far, including some of the first observing programs with JWST.

**Observations of Rocky Planet Atmospheres to Date**

A handful of rocky planets (~ 10) have been observed with transmission spectroscopy (Wordsworth & Kreidberg 2022), most of them in short period orbits around M-dwarf stars. These planets have typically shown "featureless" spectra, lacking any strong signal from candidate gases like $H_2O$ or $CO_2$ (e.g. de Wit et al. 2016, 2018; Diamond-Lowe et al. 2018, 2020; Libby-Roberts et al. 2022; Lustig-Yaeger et al. 2023); featureless spectra have also been seen for some mini-Neptunes, such as GJ 1214 b, as well (e.g. Bean et al. 2010). Although seemingly disappointing, the observed featureless spectra allow cloud-free $H_2/He_2$ atmospheres to be ruled out, meaning these rocky planets likely have either no atmosphere or higher mean molecular weight atmospheres. An atmosphere dominated by high molecular mass gases, like $H_2O$ or $CO_2$, which are the dominant products of volcanic outgassing on the modern Earth (e.g. Gaillard et al. 2021), is gravitationally bound closer to the planet's surface than a low mean molecular mass atmosphere (e.g. one dominated by $H_2$ or $He_2$). With less atmospheric cross section for starlight to pass through, high mean molecular mass atmospheres produce smaller signals than, e.g., $H_2$ or $He_2$ dominated atmospheres. The cloud-free caveat is also important, however, as another possibility for the featureless spectra is clouds masking any signal from atmospheric gases. The possibility of clouds then provides a note of caution for the prospects of observing higher mean molecular weight atmospheres with next generation telescopes; even with highly sensitive observing capabilities, clouds might mask our ability to determine the makeup of such atmospheres or search for biosignature gases (e.g. Komacek et al. 2020).

A few rocky planets, LHS 3844 b (Kreidberg et al. 2019), GJ 1252 b (Crossfield et al. 2022), and now with JWST Trappist-1 b (Greene et al. 2023) and Trappist-1 c (Zieba et al. 2023), have been observed in thermal emission. These planets all show strong dayside to nightside temperature contrasts consistent with either very thin atmospheres or airless rocky bodies. While the finding of presumably atmosphereless planets may also seem disappointing, determining which planets do and do not have atmospheres is a fundamental question for planetary evolution. Zahnle & Catling (2017) dub the boundary between airless planets and planets that retain atmospheres as the "cosmic shoreline." Zahnle & Catling (2017) empirically fit this boundary to solar system bodies with simple power-law relationships. The relationships are formulated a few different ways, which may be applicable for different stellar environments. In one case the cosmic shoreline is defined as a relationship between the incident stellar flux ($I$) and a planet's escape velocity ($v_{\rm esc}$), which is determined by its surface gravity, where $I \propto v_{\rm esc}^4$ marks the critical incident flux above which atmospheres are lost. For systems where high energy stellar radiation consisting of X-ray and extreme ultraviolet, together referred to as "XUV," radiation is dominant, then instead the integrated XUV flux ($I_{\rm XUV}$) may be more important, leading to the scaling $I_{\rm XUV} \propto v_{\rm esc}^4$ for the cosmic shoreline. The presence of absence of an atmosphere is thus assumed to be determined to first-order solely by atmospheric escape, where high escape velocity and low stellar flux means a planet can gravitationally bind its atmosphere and resist loss to space, while low escape velocity or high stellar flux means atmosphere can be easily lost.

Current work characterizing which planets do and do not have atmospheres will help test the cosmic shoreline proposed by Zahnle & Catling (2017). The candidate airless planets discussed so far experience high incident flux as they orbit very close to their host stars, much too close to lie within the liquid water habitable zone. They also likely experienced high XUV fluxes as is expected for low mass stars (Johnstone et al. 2021). In terms of incident stellar flux these four planets straddle the proposed cosmic shoreline, with GJ 1252 b and LHS 3844 b on the airless side, and Trappist-1 b and c on the side where atmospheres could potentially be retained (Figure 1). It is therefore at least somewhat surprising that Trappist-1 b and c apparently lack atmospheres today. Trappist-1 c in particular receives a comparable incident flux as modern Venus, yet instead of possessing a thick $CO_2$ dominated atmosphere, it is consistent with a bare rock or a thin $CO_2$ atmosphere. Specifically, the upper limit for a $CO_2$ poor atmosphere (10 ppm $CO_2$) is 10 bar, and for a pure $CO_2$ atmosphere the upper limit is 0.1 bar (Zieba et al. 2023).

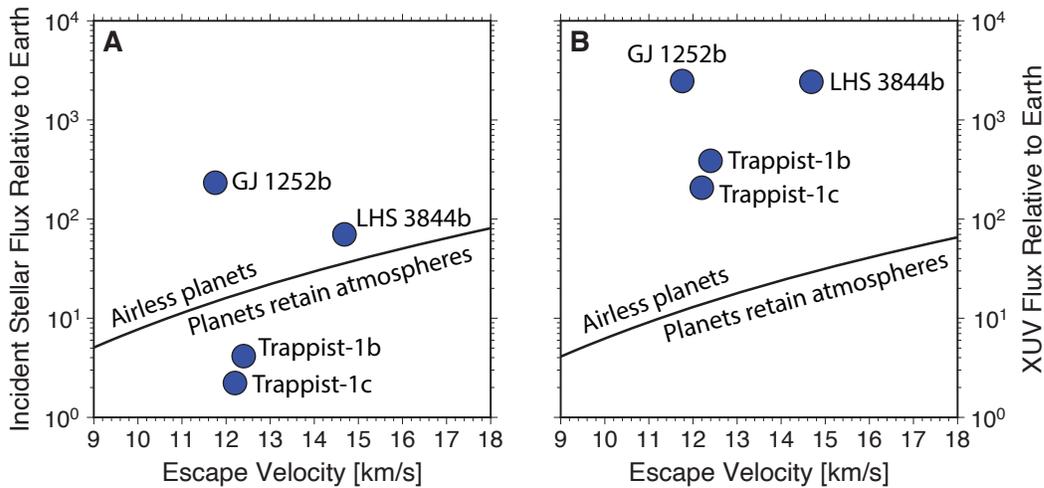

Figure 1: The cosmic shoreline proposed by Zahnle & Catling (2017) in terms of present-day stellar flux (A), $I \propto v_{esc}^4$, and integrated XUV flux (B), $I_{XUV} \propto v_{esc}^4$, with regions where planets should be airless and should retain their atmospheres labelled. The four planets interpreted to be airless based on thermal emission are plotted. Depending on the chosen formulation for the cosmic shoreline, the two Trappist planets may fall into either the regime where they would be predicted to still retain atmospheres or to be airless.

However, in terms of integrated XUV flux all four planets sit in the airless regime, indicating the potentially important role of high energy radiation for atmosphere loss, as discussed further in *Thermal vs non-thermal escape and implications for the geological processes of airless bodies*. The different predictions from the two different formulations for the cosmic shoreline also highlight how system specific factors can be important for whether planets end up retaining or losing atmospheres. In fact, even planet specific factors could be important, especially for planets falling near the cosmic shoreline, where secondary effects can tip the balance between losing or retaining an atmosphere. Such secondary effects could include additional escape processes not considered in the Zahnle & Catling (2017) model, or geological processes and characteristics of planets, like their abundance of volatiles that can form atmospheric gases or their long-term rates of volcanism and outgassing. Ultimately then a full accounting of different atmospheric sources and sinks is needed to study how airless planets come to lose their atmospheres, which can be accomplished using detailed models of planetary evolution (e.g. Kane et al. 2020; Crossfield et al. 2022; Krissansen-Totton & Fortney 2022; Krissansen-Totton 2023; Teixeira et al. 2023). Such an approach further allows constraints to be placed on airless planet's atmospheric sources and sinks, with at least some of these constraints tying back to exoplanet geology. To discuss how this can be accomplished it is necessary to first briefly review the major sources and sinks of atmospheric gases.

**Atmosphere Sources and Sinks**

The evolution of an atmosphere's size and composition is a result of the sources and sinks of gases. Therefore, knowledge of an exoplanet's atmosphere provides at least some constraint on these sources and sinks. Sources of atmospheric gasses include capture of nebula gas during planet formation, which would largely consist of $H_2$ and $He_2$, outgassing from the interior, or outgassing from impactors (Figure 2). Sinks include atmospheric loss to space by various escape processes and incorporation of atmospheric gases into surface rocks, as occurs via weathering on Earth (see e.g., Lammer et al. 2018, for a review). This section focuses on the rocky exoplanets currently inferred to lack atmospheres. These planets are all very hot, receiving stellar radiation flux higher than the runaway greenhouse limit; they thus are expected to lack liquid water oceans, meaning that weathering will not be a significant sink for atmospheric gases. These planets also will have already lost any primordial atmosphere acquired from nebula gas, and as we are looking at planets well after the formation stage, impact outgassing will be a minimal source as well. For airless rocky planets the dominant source, if any sources exist, will therefore be interior outgassing and the dominant sink will be escape to space.

On a solid surface planet, outgassing from the interior predominantly occurs by volcanism, releasing gases derived from volatiles previously stored in minerals in the crust and mantle. The discussion in this section will focus on volcanism on solid surface planets like occurs on the rocky planets in our solar system. Solid surface planets are those whose dayside temperatures fall below the melting temperature of their constituent surface rocks. Planets hotter than this will have large magma ponds or oceans on their surfaces and can possess atmospheres produced by melt-vapor equilibrium (see *Lava worlds* for a discussion of these planets). Atmospheres of lava world planets are certainly interesting, as they can help probe interior composition as discussed above, but they fall under a very different regime than outgassing produced by volcanism on a solid surface planet.

Considering volcanic outgassing as the dominant atmosphere source and loss to space as the dominant sink, then the volcanic outgassing rate could in principle be constrained for an airless body if the rate at which atmosphere could escape can be estimated or modeled. Constraints on outgassing rate in turn then constrain both volatile abundance in the planet's interior and the integrated rate of surface volcanism (see *Controls on interior outgassing*). Outgassing rate constraints will be upper bounds, and can be derived because, on an airless body the outgassing rate must be less than the atmospheric loss rate, otherwise gases would build up on the planet and an atmosphere would be present. However, both atmospheric escape and volatile outgassing are complex topics. Escape involves multiple different loss mechanisms that can dominate under different conditions (e.g. Tian 2015; Catling & Kasting 2017). The structure of the planet's atmosphere and mixing ratios of constituent gases is important in many cases, meaning that models must be able to track these properties to first order to derive reasonably accurate atmosphere loss rates. Outgassing is a result of volcanism which also occurs by a variety of different mechanisms, depends on the chemical and thermal state of a planet's interior, its style of tectonics, and conditions at which melt erupts at the surface, among other factors. In order to illustrate how knowledge of escape can be used to constrain outgassing on airless planets, it is therefore necessary to first review the basics of both outgassing and escape.

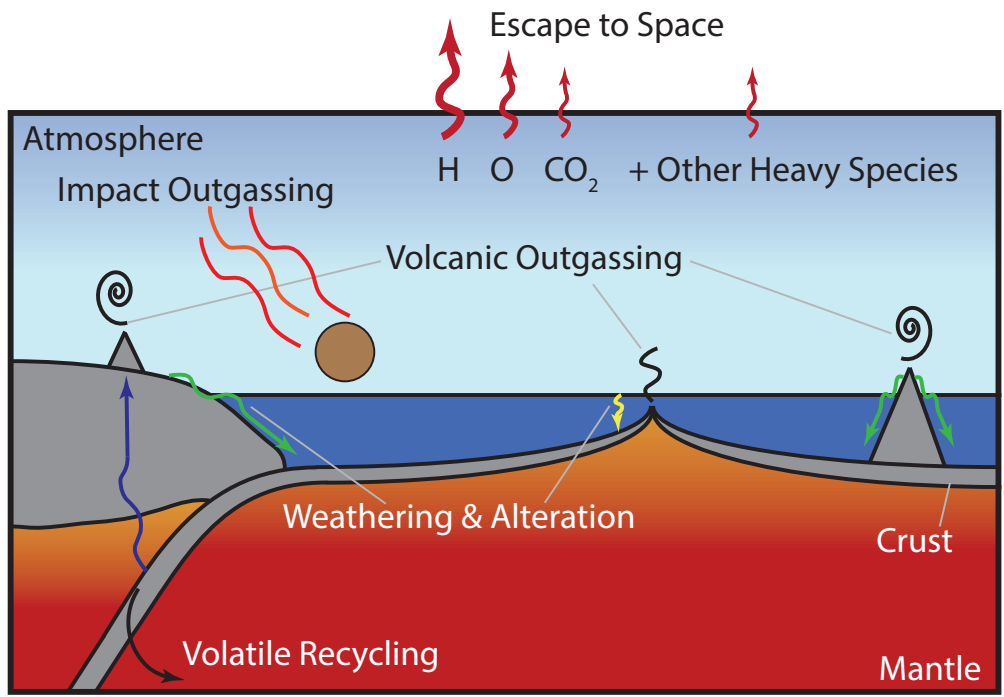

Figure 2: Schematic diagram of atmosphere sources and sinks. Adapted from Foley (2015); Foley & Driscoll (2016).

**Controls on Interior Outgassing**

Crustal and mantle rocks contain volatiles stored either in specific volatile-bearing minerals or as trace elements that take the form of defects in a mineral crystal lattice (see e.g. Hirschmann 2006, for a review of water storage in mantle minerals). When a rock melts, these volatiles are released and become incorporated into the resulting magma. As magma rises to the surface and erupts the volatiles in the magma can form into gas bubbles that eventually degas upon eruption, with the exact species and efficiency of this degassing depending on magma composition and eruption temperature and pressure conditions.

Melting is therefore an essential part of outgassing to form an atmosphere. Regions of a rocky planet can melt when the constituent rocks exceed their "solidus," or the temperature where a rock first begins to melt (see e.g. Foley et al. 2020, for an overview of this process). As rocks are mixtures of different minerals, each of which has different properties, a rock does not completely melt at one temperature. Instead, the solidus marks the temperature below which all components of the rock are solid, the "liquidus" marks the temperature above which the rock is entirely molten, and in between these temperatures the rock is partially molten, with some percentage melt and the remainder staying solid. For an Earth-like mantle composition the solidus is ≈ 1100°C at surface pressure and liquidus is ≈ 1800°C at surface pressure under dry conditions (Takahashi 1986; Katz et al. 2003). When the temperature first exceeds the solidus, minerals with the lowest melting points melt, followed by those with higher melting points as temperature increases, and finally reaching complete melting when the liquidus temperature is reached. On the modern Earth, regions experiencing melting rarely reach the liquidus, and instead only partially melt with melt fractions ranging between a few to a few 10s % (e.g. at mid-ocean ridges the mantle experiences about 10-25 % melting Spiegelman & McKenzie 1987; Kushiro 2001).

Common volatile elements that contribute to outgassing to the atmosphere on Earth are H, C, O, N, and S (e.g. Gaillard et al. 2021), and these will presumably be the primary volatile elements on rocky exoplanets as well barring planets with radically different compositions than Earth's (e.g. Lichtenberg et al. 2023). Outgassing on the modern Earth is specifically dominated by $CO_2$ and $H_2O$ (Holland 1984; Gaillard et al. 2021). Volatile elements are typically "incompatible" during melting, which means that they preferentially go into magma during melting rather than remaining behind in the residual solid. As a result, the area of mantle or crust experiencing partial melting will be devolatilized, with volatiles then becoming concentrated in the melt (see e.g., Hirth & Kohlstedt 1996; Hirschmann 2006; Korenaga 2006, for discussion of this behavior for water). This incompatible nature means that magmatism is an efficient process for removing volatiles from the interior to the surface, with some caveats discussed below.

The specific mixture of gases these volatiles make and then release to the atmosphere depends on many factors, including mantle chemistry and surface pressure (e.g. Gaillard & Scaillet 2014; Foley et al. 2020; Ortenzi et al. 2020; Lichtenberg et al. 2023). If the mantle is more oxidized, meaning more oxygen is available to form gases, then oxygen-bearing gases like $H_2O$ and $CO_2$ dominate, as is the case on the modern Earth. Under more reducing conditions gases like $H_2$, CO, or $CH_4$ dominate instead (e.g. Holland 1984; Kasting et al. 1993a; Gaillard & Scaillet 2014; Ortenzi et al. 2020). The oxidation state of the mantle also influences the partitioning behavior of mantle volatiles, with more oxidizing conditions leading to an enhanced ability of volatiles to preferentially partition into the melt during magmatism (Guimond et al. 2021).

The flux of a gas brought to the surface by volcanism can be quantified as the product of the melt eruption rate, $\dot{M}$, given here with units of $m^3 \cdot s^{-1}$, and the concentration of volatile species $i$ in the melt, $C_i$, with units of either $mol \cdot m^{-3}$ or $kg \cdot m^{-3}$. A certain fraction, $f_{\text{outgas}}$, of the gas in the magma will then outgas, with the remainder staying soluble in the melt. The outgassing flux of gas species $i$, $F_i$, is therefore (e.g. McGovern & Schubert 1989; Grott et al. 2011; Foley & Smye 2018)

$$F_i = f_{\text{outgas}} \dot{M} C_i \qquad (1)$$

Petrological models constrained by melting experiments can determine the speciation of different gases and the concentration of volatiles in the melt resulting from melting a rock of a given composition (Gaillard & Scaillet 2014). To determine the melt eruption rate, though, we must consider the tectonic setting where melting can occur.

On Earth most volcanism is caused by partial melting of the mantle, which occurs in a few different settings (Figure 3). One significant setting is at mid-ocean ridges, where mantle upwells to fill in the space left by the divergent spreading of surface plates on either side of the ridge. In this setting, upwelling mantle melts by "decompression melting," where a mantle parcel rising along an adiabat can begin to melt at shallow pressure where the solidus is lower. Most materials have lower melting temperatures at low pressure than at higher pressure, meaning a rising parcel can melt simply by moving to lower pressures rather than having its temperature rise. Decompression melting only occurs if the mantle is hot enough for the adiabat to cross the solidus before reaching the base of the planet's lithosphere, where temperatures are colder. At mid-ocean ridges, the lithosphere is very thin, allowing rising mantle to reach low pressures while still following the mantle interior adiabat, therefore facilitating melting. A different but related setting is melting associated with hot, active upwelling plumes. Plumes have an elevated temperature compared to the average mantle interior, and therefore can melt more easily and extensively during ascent. Finally, melting on Earth also occurs at subduction zones, where surface tectonic plates sink back into the interior. In these settings water incorporated into hydrated minerals in the crust and mantle of the sinking plate is released by metamorphic reactions at increasing temperature and pressure. The release of this water decreases the mantle and crustal solidus, leading to melting even at relatively low temperatures that would have been below the solidus without water present. In all these settings the rate of melt production can be calculated by estimating the flux of mantle into the region where melting occurs (where the solidus is exceeded), $\dot{m}$, and the fraction of melt in this melting region. The melt fraction, $\phi$, can be simply calculated based on the temperature within the melting region, $T$ as

$$\phi = \frac{T - T_{sol}}{T_{liq} - T_{sol}} \quad (2)$$

where $T_{sol}$ and $T_{liq}$ are the solidus and liquidus temperatures, respectively (Grott et al. 2011; Morschhauser et al. 2011; Tosi et al. 2017). The melt eruption rate, $\dot{M}$, is then

$$\dot{M} = f_{erupt} \phi \dot{m} \quad (3)$$

where $f_{erupt}$ is the fraction of melt produced that erupts at the surface.

The tectonic settings where melting occurs on Earth discussed here are in many ways tied to plate tectonics. Mid-ocean ridges, where significant volcanism occurs by decompression melting are unique to plate tectonics, as are subduction zones where melting occurs by fluid release. Upwelling mantle plumes can occur regardless of a planet's tectonic setting but are facilitated by plate tectonics. Plate tectonics leads to efficient mantle interior cooling, therefore increasing the temperature contrast at the core-mantle boundary which drives plume formation (Jellinek et al. 2002). However, plate tectonics may not be common for rocky planets if our solar system is any indication. Stagnant-lid tectonics, as present on Mars and Mercury today (Breuer & Moore 2015), represents another end-member style of tectonics where the surface is rigid and immobile, with mantle convection taking place beneath this thick "stagnant-lid" (e.g. Ogawa et al. 1991; Davaille & Jaupart 1993; Solomatov 1995; Moresi & Solomatov 1995; Stern et al. 2018). With stagnant-lid tectonics, there are no mid-ocean ridges where plates spread apart or subduction zones where plates come together. This difference in tectonics probably limits the rate and longevity of volcanism and outgassing on stagnant-lid planets but does not eliminate it for otherwise Earth-like planets (Kite et al. 2009; Grott et al. 2011; Tosi et al. 2017; Foley et al. 2020; Unterborn et al. 2022). Very large planets > 3−4 Earth masses ($M_⊕$) may have volcanism entirely suppressed due to effects of pressure discussed below, however (Noack et al. 2017; Dorn et al. 2018).

Volcanism on stagnant-lid planets can occur via processes that are analogous to those that operate on the plate-tectonic Earth, but with key differences due to the change in tectonic mode. Melting can occur in mantle upwelling zones when material crosses the solidus before reaching the base of the lithosphere, as on the plate-tectonic Earth. Upwelling can be both active, as in a mantle plume, or passive, where ambient mantle rises to compensate for mantle flowing down into the interior at focused, active downwellings (this is analogous to the style of upwelling at mid-ocean ridges on Earth) (e.g. Reese et al. 1999; Hauck & Phillips 2002; Fraeman & Korenaga 2010; Morschhauser et al. 2011; Tosi et al. 2017; Foley & Smye 2018). However, on a stagnant-lid planet the lithosphere overlying the actively convecting mantle interior is far thicker than on a plate tectonic planet, so upwelling mantle must be at much hotter temperature in order to cross the solidus before reaching the base of the lithosphere (Figure 4). This effect acts to suppress melting and volcanism but is compensated at least somewhat by stagnant-lid convection leading to higher mantle temperatures due to less efficient heat loss when compared to a plate-tectonic planet, all else equal (see Foley et al. 2020, for a review and summary of these effects). Another mechanism for volcanism on a stagnant-lid planet is foundering of the lower crust or lithosphere. This can occur by gravitational instability of regions of the lid that are denser than their surroundings due to differences in their chemical composition or phase. Foundering drives volcanism by upwelling of mantle into the space left by sinking crust or lithosphere, causing decompression melting, and potentially melting of the sinking crust itself. If the foundering crust contains volatiles, potentially due to interaction with a surface hydrosphere or atmosphere before crust was buried to mid- to lower-lithosphere depths by lava flows, then volatiles can potentially be transported back to the interior of stagnant-lid planets, or drive mantle melting by volatile release, analogous to subduction zones on the modern Earth (Elkins-Tanton et al. 2007).

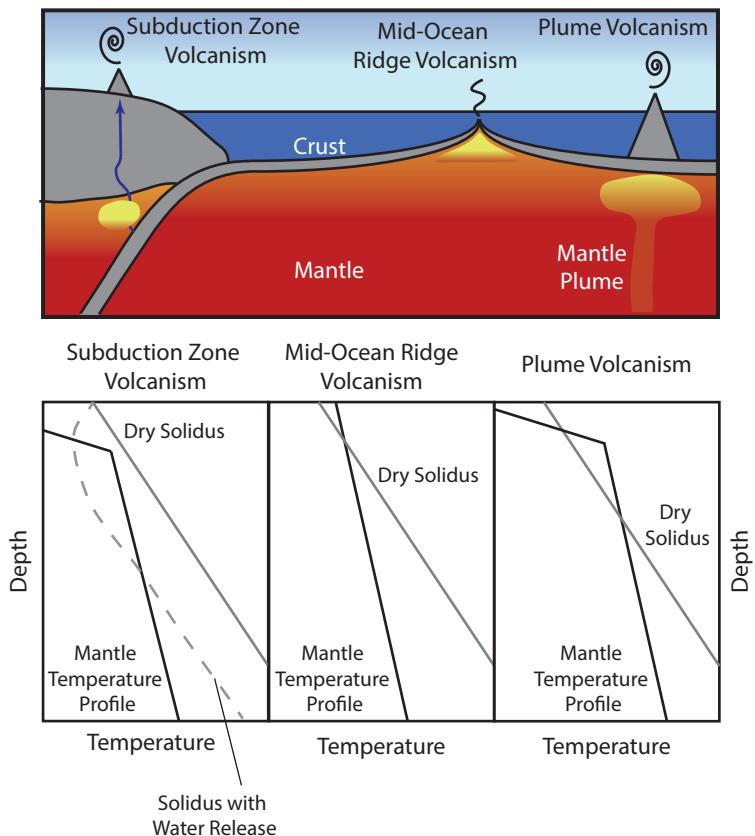

Figure 3: Schematic diagram of melting environments on the modern Earth. Bottom panels show temperature profiles with depth through the mantle in the different settings, along with the melting

curve, or solidus. In subduction zones melting occurs because release of water lowers the solidus, at mid-ocean ridges hot mantle reaches the near surface and melts by decompression, and at mantle plumes elevated temperatures cause mantle to melt.

Volatiles stored in crustal rocks can also be released directly through metamorphism, without melting the host rock, in some cases. During continental collision crust is brought to high temperatures and pressures, which are sufficient to cause metamorphic breakdown of volatile bearing minerals. These volatiles can then be incorporated into fluids and released to the atmosphere (Tajika & Matsui 1992; Sleep & Zahnle 2001). On a stagnant-lid planet, burial of crust under repeated lava flows can similarly take volatile bearing minerals to the temperature and pressure conditions where they break down and release their volatiles (Foley & Smye 2018; Höning et al. 2019). If released volatiles can percolate to the surface through cracks and pore space, then this metamorphic breakdown contributes an additional source of atmospheric gases.

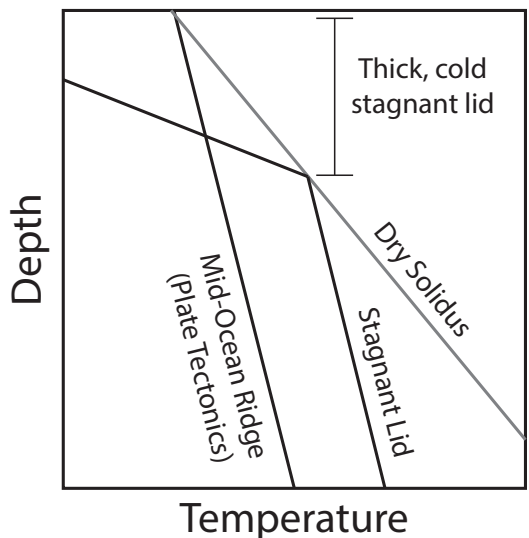

Figure 4: Schematic diagram showing the coldest temperature profile that can still intersect the solidus for a stagnant-lid planet and for a plate tectonic planet at a mid-ocean ridge setting. The thick, cold stagnant lid requires hotter temperatures in the mantle for melting to occur.

Volcanism and subsequent outgassing of volatiles to form an atmosphere should be a common process on rocky planets, regardless of tectonic regime (see also Shorttle and Sossi (20XX, this volume); Dasgupta et al (20XX, this volume) for a more detailed discussion). However, there are important caveats. On large ($> 3 - 4$ $M_\oplus$) stagnant-lid planets melt production and eruption may be inhibited due to the effects of increased pressure in the interior. With increasing planet mass, the pressure gradient in the mantle increases. As a result, the temperature required to cause melting of the mantle beneath a lithosphere of a given thickness is higher on a larger planet than a smaller one, which tends to suppress melting. Moreover, even if melt forms, above pressures of $\sim 10$ GPa silicate melts can be denser than solids, meaning melt would sink rather than rise; volcanism would not occur in this case. Both effects have been argued to suppress volcanism on large stagnant-lid exoplanets (Noack et al. 2017; Dorn et al. 2018). However, the result is sensitive to assumptions about the viscosity structure of the mantle, which controls the convective vigor that in turn sets the thickness of the stagnant lid. Larger planets retain heat better than smaller planets, leading to higher temperatures which also tend to promote thinner lithospheres, both effects acting to promote volcanism. As a result, volcanism may still be active even on large rocky exoplanets in a stagnant-lid regime, at least with sufficient heat production by radioactive decay or tidal heating (Unterborn et al. 2022). Miyazaki & Korenaga (2022) argue that partitioning of volatiles like $H_2O$ and $CO_2$ during magma ocean solidification will also influence later outgassing, with planets that retain water in their interiors likely to fall into stagnant lid states and experience limited outgassing of water. These planets in their models would still develop $CO_2$ atmospheres, however.

Another important caveat is that volcanic outgassing is suppressed at higher surface pressures, as volatiles are generally more soluble in melt with increasing pressure; this means volatiles erupted at high surface pressure will remain dissolved in lava rather than being released as gases. One result of this solubility effect is that planets with thick atmospheres or water oceans may experience limited outgassing (Krissansen-Totton et al. 2021). Suppression of outgassing beneath a thick atmosphere or ocean is important for considering the potential habitability and evolution of such exoplanets, but not relevant to the discussion in this section focused on airless bodies, and what they tell us about rocky planet interiors.

**Thermal vs Non-Thermal Escape and Implications for the Geological Processes of Airless Bodies**

Escape can remove a substantial mass of atmospheric gases from a planet, depending on atmosphere composition, planet size, and stellar radiation, among other factors. The gap between rocky super-Earths and sub-Neptunes with thick $H_2$ and $He_2$ dominated atmospheres may be explained by atmospheric loss on the rocky planets, meaning the loss of up to a few percent of the planet's mass by escape (e.g. Owen & Wu 2013; Ginzburg et al. 2018). Atmospheric escape can occur by two primary groups of mechanisms, thermal escape and non-thermal escape (Catling & Kasting 2017). Thermal escape encompasses mechanisms by which stellar radiation heats an atmosphere providing enough energy for molecules to escape, and non-thermal escape involves chemical reactions or ionic interactions providing the necessary energy for molecules to escape. Non-thermal escape is often associated with stripping of atmospheric molecules by interactions with stellar winds, and models of stellar wind-atmosphere interactions can provide first order estimates of non-thermal escape rates for exoplanets (e.g. Dong et al. 2017; Garraffo et al. 2017; Dong et al. 2018). Impacts can also cause significant atmospheric stripping (Schlichting et al. 2015). Impact fluxes are high during and just after planet formation, so impact erosion can be important for helping remove any primordial atmosphere or atmosphere formed by very early interior outgassing but will be a minor loss process for the majority of planets' lifetimes.

Both thermal escape and non-thermal escape via stellar wind stripping can be significant sources for atmospheric loss, especially for planets around M-dwarf stars like the handful of presumably airless planets considered here. M-dwarf stars emit high energy XUV radiation at an elevated rate for longer than higher mass stars (e.g. Johnstone et al. 2021). This high XUV flux can drive rapid thermal escape of hydrogen, especially when the planet is in a runaway greenhouse state where water vapor is abundant in the upper atmosphere (Luger & Barnes 2015). The airless rocky exoplanets all receive stellar radiation flux above the runaway greenhouse limit. Thus, if water vapor was present in the atmosphere or being outgassed from the interior, it would remain in vapor form and could well mix into the stratosphere for rapid, XUV driven escape. This rapid thermal escape can also potentially drag away heavier gases like $O_2$ or $CO_2$, leading to significant loss of all major atmospheric gases (Zahnle & Kasting 1986; Odert et al. 2018; Krissansen-Totton

& Fortney 2022; Krissansen-Totton 2023; Zahnle & Kasting 2023). Higher levels of stellar activity also enhance stellar wind stripping, so planets around M-dwarf stars will also experience elevated rates of non-thermal escape (e.g. Dong et al. 2018).

We can explore how these different escape mechanisms would constrain present-day outgassing rates on LHS 3844 b and Trappist-1 b and c by recognizing that present-day outgassing must be less than atmosphere loss, otherwise an atmosphere would be accumulating on these planets. In other words, estimating present day atmospheric escape rates then also gives the maximum allowable present day outgassing rate (Figure 5). For illustrative purposes only non-thermal escape due to stellar wind stripping is considered. Three-dimensional magneto-hydro-dynamic models have been used to infer mass loss rates via stellar wind stripping of $\sim 170$ kg·s$^{-1}$ and $\sim 60$ kg·s$^{-1}$ for Trappist-1 b and c, respectively (Dong et al. 2018). Kreidberg et al. (2019) estimated the present-day stellar stripping rate for LHS 3844 b at 30-300 kg·s$^{-1}$ based on simulations of stellar winds for Proxima Cenatauri (same stellar class as LHS 3844) from Dong et al. (2017). Garraffo et al. (2017) argue for higher stripping rates by about an order of magnitude for the Trappist-1 system. All told estimates range from order of $10 - 1000$ kg·s$^{-1}$ mass loss rates.

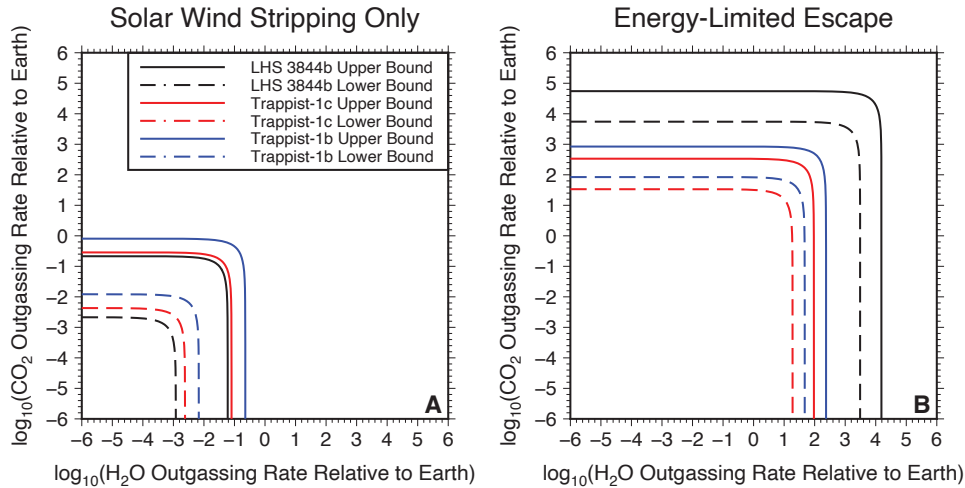

Figure 5: Contour lines representing the combined maximum allowable present day outgassing rates of $CO_2$ and $H_2O$, relative to the Earth's present day outgassing rates, for LHS 3844 b, Trappist-1 b, and Trappist-1 c. Estimates based on mass loss rates from solar wind stripping only (A), and from energy-limited escape (B) are shown. Upper bounds on the maximum allowable outgassing rates are calculated using the upper limits on mass loss rates and lower limits on Earth's present-day $CO_2$ and $H_2O$ outgassing rates, while lower bounds use the lower limits on mass loss rates and upper limits on Earth's outgassing rates.

However, compared to outgassing on the modern Earth, these atmospheric stripping rates from stellar winds are small (Figure 5A). Outgassing on Earth is dominated by $CO_2$ and $H_2O$. The rate of $CO_2$ outgassing is estimated at $\sim 10^{12}-10^{13}$ mol·yr$^{-1}$ or $\sim 10^3-10^4$ kg·s$^{-1}$ (Marty & Tolstikhin 1998), while outgassing of $H_2O$ is estimated at $\sim 1 - 4.5 \times 10^{13}$ mol·yr$^{-1}$, or $\approx 5\times10^3-2.5\times10^4$ kg·s$^{-1}$ (e.g. Parai & Mukhopadhyay 2012; Peslier et al. 2017). Summing these estimates outgassing releases atmospheric gases at a rate of $\approx 6 \times 10^3 - 3.15 \times 10^4$ kg·s$^{-1}$ on the modern Earth, ranging from a factor of a few to up to 3 orders of magnitude larger than the atmospheric stripping rates, even the highest estimate for Trappist-1 b. Even removing outgassing at subduction zones, which primarily reflects release of $CO_2$ and $H_2O$ incorporated onto subducting plates by weathering, likely to be absent on the planets considered here as they are too hot for liquid water oceans, the result remains the same. $CO_2$ degassing decreases by about 1/3, and $H_2O$ degassing is decreased to $\sim 1-1.5\times10^{13}$ mol·yr$^{-1}$. These reduced outgassing flux ranges still lead to total rates of $\sim 5.6\times10^3 -1.5\times10^4$ kg·s$^{-1}$, much larger than the range of stellar wind stripping rates.

Thermal escape rates can be much higher, leading to a much larger range of allowable present day outgassing rates (Figure 5B). Estimating thermal escape requires knowledge of the mixing ratios of different atmospheric species as well atmospheric structure, and therefore requires a model coupling outgassing to atmosphere evolution and escape; a simple general estimate cannot be provided. However, based on measured or modeled XUV fluxes, an upper bound can be estimated by assuming energy-limited escape, where all the XUV energy absorbed by the atmosphere contributes to causing escape (e.g. equating the absorbed XUV flux with the energy carried by escaping molecules, (Catling & Kasting 2017)). Diamond-Lowe et al. (2021) estimated the modern-day energy-limited escape flux for LHS 3844 b as $\sim 7.7 \times 10^7$ kg·s$^{-1}$ and Crossfield et al. (2022) estimated a similar value, $\sim 2 \times 10^7$ kg·s$^{-1}$, for the airless exoplanet GJ 1252 b. Wheatley et al. (2017) estimated energy-limited escape rates of $1.18 \times 10^6$ kg·s$^{-1}$ for Trappist-1b, and $4.7 \times 10^5$ kg·s$^{-1}$ for Trappist-1c. Clearly under this energy-limited escape regime atmosphere loss is significantly higher than when considering only stellar wind stripping. Energy-limited escape strictly only applies to H, but rapid H escape can also drag away heavier species like $CO_2$ (Krissansen-Totton 2023), so it is reasonable for at least a first order estimate to consider this loss rate applying to the whole atmosphere. Outgassing rates up to 4-5 orders of magnitude higher than modern Earth's would be needed to still buildup an atmosphere for LHS 3844 b in the face of such high atmosphere loss rates. For the Trappist planets outgassing rates up to 2-3 orders of magnitude larger than modern Earth's would still keep the planets airless (Figure 5B). It should be noted that these energy-limited atmosphere loss rates are upper bounds, and real rates could be much lower if H is not abundant in the upper atmosphere or significant a fraction of the XUV energy is not available to drive thermal escape.

The simple calculations presented here show that for an airless rocky planet, the lower the present-day rate of atmospheric loss (if loss rates can be well estimated), the tighter the constraint on present-day outgassing. If energy-limited escape is a first-order accurate estimate of loss rates on the planets considered, then even outgassing much more rapid than occurs on the modern Earth would still result in airless planets. Thus, a wide range of outgassing rates would be compatible with the lack of atmosphere, and the constraint is weaker. However, further context for cases where atmosphere mass loss rates are high can be provided by considering how large outgassing rates could conceivably get on rocky planets. Rocky planets generally cool over time as they lose heat from formation and radioactive heat sources decay (e.g. Davies 2007; Breuer & Moore 2015; Foley et al. 2020). As a result, early in a planet's history volcanism, and hence outgassing, rates are higher due to more extensive melt production (i.e. both larger melt fractions, $\phi$, and fluxes of mantle material into the melting region, $\dot{m}$, see Eq. 3). Outgassing rates of both $H_2O$ and $CO_2$ can easily exceed 2-3 orders of magnitude larger than present day rates during a planet's early evolution when heat sources were larger (e.g. McGovern & Schubert 1989; Tajika & Matsui 1992; Sleep & Zahnle 2001; Crowley et al. 2011; Driscoll & Bercovici 2013; Foley & Smye 2018). Therefore, even the high mass loss rates for energy-limited escape provide a meaningful constraint for present day outgassing on Trappist-1 b and c, as some geologically plausible outgassing rates can be excluded. In fact, both Trappist-1b and c likely experience significant tidal heating, estimated to provide a heat flux of $\approx 1$ W·m$^{-2}$ and $\approx 0.6$ W·m$^{-2}$ for b and c, respectively (Dobos et al. 2019). These heat fluxes are huge compared to the Earth; Earth's heat flux through oceanic plates is $\approx 0.1$ W·m$^{-2}$ (Jaupart et al. 2015). Tidal heating is therefore sufficient to drive extensive volcanism, and if interior volatile abundances were Earth-like, extensive outgassing, potentially large enough to outpace atmospheric escape. The lack of an atmosphere on Trappist-1 b and c may therefore require that the planets are volatile poor today. Indeed Krissansen-Totton & Fortney (2022) argues that even higher rates of escape during the early evolution of the Trappist-1 system can thoroughly deplete planets b and c of volatiles, such that even with active volcanism today outgassing rates would be muted. For the energy-limited escape rates estimated for LHS 3844 b, even a planet significantly more enriched in volatiles than the Earth would still have its volatiles stripped away once outgassed, as found by Crossfield et al. (2022) for GJ 1252 b, a planet with similar energy-limited escape rates.

Although thermal escape is the more efficient, and therefore probably dominant escape process on the airless planets discussed here, it is still illustrative to consider the implications for these planets if stellar wind stripping, with its relatively lower mass loss rates, were the only major escape process. Future observations of additional planets may reveal airless bodies where total escape rates, even including thermal escape, are comparable to the stellar wind stripping rates discussed here. Such cases provide the tightest constraints on present day outgassing rates, and thus implications for planetary volcanism rates or volatile budgets. As an example, take the lower bounds on the maximum allowable present day outgassing rates based solely on solar wind stripping for LHS 3844 b as the total atmospheric loss rate. In this case, outgassing rates would need to be 2-3 orders of magnitude lower than the present-day Earth's for LHS 3844 b to lack an atmosphere. This can be achieved either through much lower rates of interior melting and subsequent volcanism, much lower interior volatile abundances, or a combination of the two. For LHS 3844 b either option would be plausible. As LHS 3844 b is the only known planet in its system, planet-planet interactions are not available to pump up the orbital eccentricity and drive high rates of tidal heating like in the Trappist system. Given the age of the system (7.8± 1.6 Gyr), it is plausible that LHS 3844 b has cooled to the point where volcanism has ceased (Kane et al. 2020; Unterborn et al. 2022), meaning no present day outgassing even with volatile stores in the interior. A volatile poor interior, either due to loss of volatiles early in the planet's history or a volatile poor formation, is also always a possibility for any planet where present-day outgassing rates can be inferred to be low. In fact, Kane et al. (2020) argued that with only stellar wind stripping as an atmospheric loss process for LHS 3844 b, the planet would need to have formed volatile poor, as outgassing earlier in the planet's history would build up an atmosphere too large to be stripped away by the present day with an Earth-like volatile budget. For Trappist-1 b and c, significant tidal heating means volcanism is likely. As a result, if stellar wind stripping were the only loss process, then Trappist-1 b and c's interiors would need to be volatile depleted to keep outgassing rates low enough to be airless (Teixeira et al. 2023).

**Implications for Outgassing History and Future Directions**

The discussion in the previous section focused on present day outgassing and atmospheric loss rates. However, an airless planet is not just a snapshot of the present-day state, but also a function of the evolutionary history of the planet up to that point. For example, a planet that is airless today could be a result of having lost its atmosphere and planetary volatile store during an earlier period of rapid escape, as may be the case for Trappist-1 b and c. The best avenue for constraining the geology and geologic history of planets currently observed to be airless is through models coupling interior outgassing, atmosphere evolution, and escape using broad sweeps through the relevant parameter space, as in e.g. Krissansen-Totton & Fortney (2022); Krissansen-Totton (2023). These papers argue that Trappist-1 b and c experience rapid XUV driven thermal escape during their early evolution that left them volatile depleted and airless. Trappist-1 b and c receive stellar radiation fluxes higher than the runaway greenhouse limit, leading to high mixing ratios of H in the upper atmosphere and hence rapid escape. Meanwhile, Trappist-1 e and f could still host atmospheres today because they formed further out and receive radiation less than the runaway greenhouse limit. As a result, water can condense and H mixing ratios in the upper atmosphere are low, limiting escape. It is important that early escape not only remove atmosphere but fully deplete the interior volatile stores on Trappist-1 b and c as well for them to still be airless today; without such complete volatile depletion later volcanism could re-form atmospheres (Teixeira et al. 2023). However, with complete volatile depletion then even with significant later volcanism there will be little to outgas, limiting later atmosphere development. Observations from JWST will hopefully soon determine whether Trappist-1 e or f have atmospheres. If they are instead airless, though, then much tighter constraints on these planets' geologies would be placed, as they would likely need to be either volatile poor or have muted volcanism in order to not still have observable atmospheres (Krissansen-Totton 2023).

The scenario of hydrodynamic loss of H dragging away heavier species like $CO_2$ to leave behind a volatile depleted and airless planet may be complicated by additional factors, though, like the planet's composition and interactions between the atmosphere and planetary interior. These complicating factors, however, allow tighter constraints to be placed on planetary evolution for airless planets, as scenarios where even high XUV fluxes still leave an atmosphere behind, or volatiles in the interior that can outgas to form an atmosphere, after the star settles on the main sequence can then be ruled out. If rocky planets are in a magma ocean phase when experiencing high stellar fluxes that lead to a runaway greenhouse climate, then the abundance of gases in the atmosphere will be determined by equilibrium with the rapidly convecting magma. Sossi et al. (2020) and Bower et al. (2022) argue that the high solubility of $H_2O$ in magma means that magma ocean atmospheres are dominated by CO or $CO_2$, depending on oxidation state, rather than steam. Steam atmospheres only form at the end of magma ocean solidification and only if magma-atmosphere equilibrium can be maintained until the end of solidification. If a planet in a runaway greenhouse state has a relatively dry, CO or $CO_2$ dominated atmosphere, then H escape could be slower as H will have to diffuse through the dominant species to reach the upper atmosphere. This in turn can slow loss of the heavier species dragged away by H as well, leading to more scenarios where planets can retain atmospheres. More work will be needed to determine just where such scenarios occur and what can then be constrained about planets that are inferred to be airless. Planetary volatile budget and oxidation state will likely be important as well, and therefore some ranges of these parameters may be able to be ruled out for airless planets. As explained in *Controls on Interior Outgassing* above, mantle oxidation state controls whether outgassing is dominated by reduced species like $H_2$, CO, or $CH_4$, or oxidized species like $H_2O$ and $CO_2$. Moreover, bulk interior volatile inventories will further dictate whether e.g. H-bearing, C-bearing, or even more exotic gases like S-bearing species dominate released gases. As high H mixing ratios in the upper atmosphere are needed for rapid escape, some planets whose compositions lead to atmospheres dominated by other gases may be able to retain atmospheres more readily. Therefore, even in environments that favor high escape rates, some combinations of planetary volatile inventories and oxidation states might still leave atmospheres behind, and such parameter combinations can thus be ruled out for airless bodies.

Overall, airless exoplanets actually present an excellent opportunity to learn about geologic characteristics and geologic history of these planets. Building a large sample of airless rocky planets will not only help define the cosmic shoreline, but also potentially help test models of interior outgassing discussed in *Controls on Interior Outgassing*. For example, Noack et al. (2017); Dorn et al. (2018) propose that high mass rocky exoplanets in a stagnant-lid regime will experience limited outgassing. Determining whether planets operate in stagnant-lid or plate tectonic regimes will be difficult, and potentially implausible, as there are no known signatures of tectonic regime that can be realistically observed with current and upcoming missions. However, if a large database of airless exoplanets is built up, then seeing a preference for high mass planets to be airless would lend credence to the models of Noack et al. (2017) and Dorn et al. (2018). Moreover, a large sample of airless exoplanets could also test models of the longevity of volcanic activity (e.g. Unterborn et al. 2022). The age distribution of airless rocky planets, combined with models or constraints on escape rates, could help estimate when rocky planet outgassing wanes. The seemingly disappointing discovery of airless rocky exoplanets thus actually provides an excellent opportunity to study exoplanet geology.

# FUTURE DIRECTIONS: TESTING THE CARBONATE-SILICATE CYCLE FEEDBACK

This chapter so far has focused on current observations or those feasible in the near-term with current resources like JWST. However, the ability to characterize rocky exoplanets will improve greatly in the medium to long-term through new direct imaging missions like the proposed HWO or mid-IR LIFE mission. These new capabilities will open up a wide range of new questions that can be asked and models for exoplanet dynamics and evolution that can be tested. One of the most important hypotheses that can potentially be tested is the carbonate-silicate cycle weathering feedback, thought to be essential for sustaining the long-term habitability of Earth (e.g. Walker et al. 1981; Kasting & Catling 2003; Abbot et al. 2012; Foley & Driscoll 2016).

The carbonate-silicate cycle is the long-term cycling of carbon between Earth's interior, atmosphere, ocean, and crust, which acts to stabilize climate against changes in solar luminosity or internal perturbations like varying $CO_2$ outgassing rates (Berner 2004; Foley & Driscoll 2016). $CO_2$ is released to the atmosphere by outgassing, primarily from volcanism. This $CO_2$ is then removed by silicate weathering, the breaking down of silicate rocks on the surface by acidic rainwater to form ions that flow through groundwater to rivers and eventually the oceans. In the oceans carbonate minerals and $SiO_2$ precipitate, locking $CO_2$ away in seafloor rocks. Eventually this geologically stored $CO_2$ reaches a subduction zone, where a fraction of the $CO_2$ is liberated from the sinking plate by metamorphic reactions and returns to the atmosphere, while the remainder is subducted into the mantle interior, closing the cycle (Figure 6). The rate of silicate weathering, which removes $CO_2$ from the atmosphere, depends on temperature, soil pH, and the rate at which water flows through the subsurface, leading to a negative climate feedback (Kump et al. 2000; Brantley & Olsen 2014): at higher temperatures and atmospheric $CO_2$ levels, weathering rates increase acting to lower $CO_2$ and cool the climate, while at low temperatures weathering rates decrease allowing outgassing to build $CO_2$ up in the atmosphere and warm the climate.

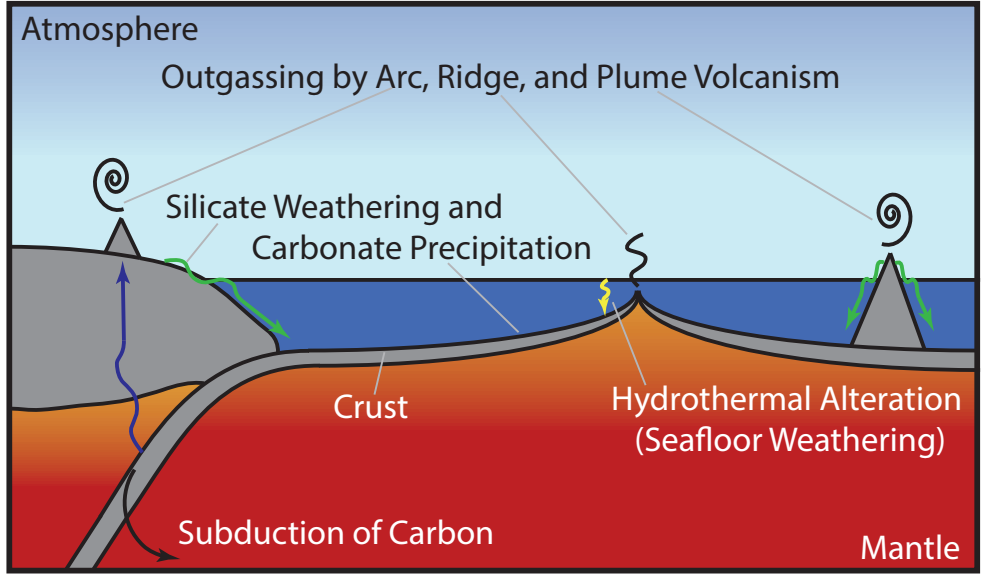

Figure 6: Schematic diagram of the carbonate-silicate cycle as it operates on the modern-day, plate tectonic Earth. Adapted from Foley (2015); Foley & Driscoll (2016).

The negative feedback mechanism inherent in the carbonate-silicate cycle acts to keep the rate of atmospheric $CO_2$ removal in balance with the rate of $CO_2$ release to the atmosphere by outgassing (Berner & Caldeira 1997), so the system can be well understood by looking at climate states that result from a steady state between weathering and outgassing (Walker et al. 1981; Berner et al. 1983; Sleep & Zahnle 2001). A simple form of the rate at which weathering on land removes $CO_2$ from the atmosphere ($F_w$) can be given by (e.g. Foley & Driscoll 2016)

$$F_w = F_w^* \exp\left[\frac{E}{R_g}\left(\frac{1}{T^*} - \frac{1}{T_s}\right)\right]\left(\frac{P}{P^*}\right)^\beta \quad (4)$$

where $F_w \approx 6 \times 10^{12}$ mol·yr$^{-1}$ is the present-day silicate weathering flux, $E$ is an activation energy for silicate weathering on land, $R_g$ is the universal gas constant, $T_s$ is the surface temperature, $T^* \approx 288$ K is the preindustrial surface temperature of Earth, $P$ is the partial pressure of atmospheric $CO_2$, $P^* \approx 30$ Pa is the preindustrial atmospheric $CO_2$ on Earth, and $\beta$ is a constant. Some models explicitly include a term for runoff, which is related to the precipitation rate. However, precipitation scales with climate, as warmer climates tend towards higher precipitation rates, so runoff can essentially be grouped into the temperature and atmospheric $CO_2$ dependent terms (e.g. Berner 2004). Key parameters like $E$ and $\beta$ are inferred from lab experiments and models (e.g. Walker et al. 1981; Brantley & Olsen 2014; Krissansen-Totton & Catling 2017) but are still uncertain. In fact, the formulation itself for weathering flux is uncertain, as Eq. 4 does not take into account other potentially important factors like uplift and erosion (e.g. Kump & Arthur 1997; Maher & Chamberlain 2014), limits to weathering imposed by the formation of thick soils (West et al. 2005; West 2012; Foley 2015; Kump 2018), weathering of ocean crust on the seafloor (Krissansen-Totton & Catling 2017; Krissansen-Totton et al. 2018; Coogan & Gillis 2013; Coogan & Dosso 2015), or effects of continental versus ocean coverage (Abbot et al. 2012; Hayworth & Foley 2020).

Nevertheless, Eq. 4 still helps illustrate how the carbonate-silicate cycle operates to regulate climate, and the trends in atmospheric $CO_2$ that would be expected for planets with this cycle in operation that could be looked for with next generation space telescopes. Specifically, planets within the habitable zone should show an increase in atmospheric $CO_2$ moving from those near the inner edge of the habitable zone towards those near the outer edge (Bean et al. 2017; Turbet 2020; Lehmer et al. 2020; Lustig-Yaeger et al. 2022). At the simplest level, this is due to the need for higher concentrations of atmospheric $CO_2$ to keep the climate warm when received stellar radiation is low, as found near the outer edge of the habitable zone, and less atmospheric $CO_2$ at high incident stellar flux as found near the inner edge. Bean et al. (2017) and Turbet (2020) assumed a constant temperature for planets through the habitable zone in order to calculate how atmospheric $CO_2$ would vary as a function of orbital distance or incident stellar flux. Clearly planets further from their host star would need higher atmospheric $CO_2$ levels to maintain the same temperature as planets closer in, all else equal.

Although it does not drastically change the overall trend of atmospheric $CO_2$ versus orbital distance or incident stellar flux, Lehmer et al. (2020) pointed out that planets within the habitable zone with an active carbonate-silicate cycle will not sit at the same surface temperature, all other factors equal, but instead show decreasing temperatures with increasing orbital distance. Atmospheric $CO_2$ still increases with orbital distance, but with a different functional form than if one assumes constant surface temperatures prevail throughout the habitable zone. The reason for this can be seen from the following thought experiment: Consider two planets, one lying within the habitable zone but close to the inner edge (planet b), and a second lying further out, closer to the outer edge of the habitable zone (planet c). The two planets have the same geologic features including the same rate of $CO_2$ outgassing. As the carbonate-silicate cycle acts to keep the $CO_2$ weathering flux in balance with $CO_2$ outgassing, then both planets will have the same weathering flux, $F^0_w$. Planet c receives less incident stellar flux than b, meaning that c would be colder if atmospheric $CO_2$ levels were the same. This cooler surface temperature would result in a lower a weathering flux, which the carbonate-silicate cycle would correct for by allowing more $CO_2$ to build up in the atmosphere, until the weathering flux for planets b and c are again equal at $F^0_w$. Equating the two planets' weathering fluxes and for the sake of generality assuming that both their atmospheric $CO_2$ levels and surface temperatures are different gives:

$$\frac{1}{T_c} - \frac{1}{T_b} = \frac{R}{E} \ln \left(\frac{P_c}{P_b}\right)^\beta \tag{5}$$

where subscripts $b$ and $c$ denote planets b and c, respectively. With $P_c > P_b$, as it should be based on orbital distance and the carbonate-silicate cycle feedback, then the right-hand side of Eq. 5 is positive. For the left-hand side to also be positive, $T_c < T_b$. Eq. 5 also shows that if $\beta = 0$, that is there is no direct influence of atmospheric $CO_2$ on weathering, then $T_c = T_b$. Or in other words, with $\beta = 0$ the only way for two planets at two different orbital distances to have the same weathering flux, $F^0_w$, is for them to have the same temperature; $CO_2$ would adjust to whatever value was needed to accomplish this in such a case.

Atmospheric $CO_2$ and surface temperature as a function of incident stellar flux is given in Figure 7, using Eq. 4 and the simple climate parameterization from Abbot et al. (2012). The calculation assumes $E = 30$ kJ·mol$^{-1}$ and $\beta = 0.33$ (Krissansen-Totton & Catling 2017), and shows the expected increase in atmospheric $CO_2$ with decreasing incident stellar flux through the habitable zone. However, as Lehmer et al. (2020) demonstrates, a sample of real exoplanets would be likely to have significant variations in their geological features, which will lead to significant variability in quantities like $CO_2$ outgassing rate, activation energy for weathering, $\beta$, or other factors. These differences in geological characteristics from planet to planet will cause a significant amount of scatter about the trend of increasing atmospheric $CO_2$ with decreasing flux, meaning that a larger number of planets would need to be observed to detect the trend.

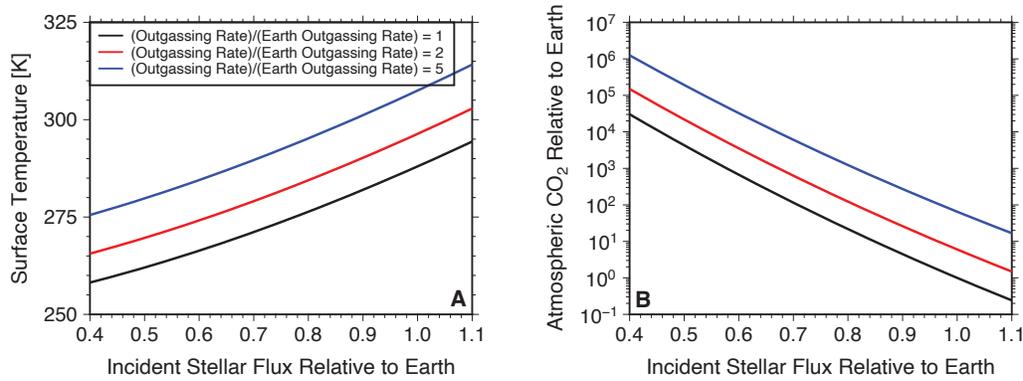

Figure 7: Surface temperature (A) and atmospheric $CO_2$ (B) as a function of relative stellar flux received resulting from a steady state between silicate weathering and $CO_2$ outgassing. Models presented assume $E = 30$ kJ·mol$^{-1}$, $\beta = 0.33$, and $CO_2$ outgassing rates relative to the modern-day Earth value as indicated by the legend.

A simplified version of the analysis from Lehmer et al. (2020) is presented here (Figure 8). 1000 calculations, each representing a hypothetical planet, are run, sampled from uniform distributions of the parameters $E$ from 15-30 kJ·mol$^{-1}$, $\beta$ from 0.24-0.44 (both as estimated from Krissansen-Totton & Catling 2017), and the $CO_2$ outgassing rate relative to the Earth in logspace ranging from 0.3 to 3 times the nominal present day Earth's value of $\approx 6 \times 10^{12}$ mol·yr$^{-1}$. The outgassing rate variations in particular are small compared to range of outgassing rates possible over a planet's lifetime (see *Thermal vs non-thermal escape and implications for the geological processes of airless bodies*). However, even with this underestimate of the true variability, modeled planets scatter by orders of magnitude in atmospheric $CO_2$ around the expected trend of increasing $CO_2$ with decreasing incident stellar flux (Figure 8A). From the complete set of 1000 modeled planets, random subsamples of planets, ranging from 5-100, are drawn and fit to a linear relationship between $\log_{10}(P)$ and incident stellar flux. The slope in this space should be negative reflecting the decrease in atmospheric $CO_2$ with increasing flux. For samples of planets less than

~ 10, the slope found from randomly sampling planets shows a huge variability, from very strongly negative slopes, much lower than the "true" slope determined from a fit to the entire set of 1000 modeled planets, to even obtaining positive slopes, the reverse of the expected trend (Figure 8B). This simple exercise shows that > 10s of planets will need to be observed to confidently detect the expected trend of decreasing $CO_2$ with increasing flux, as shown with a more comprehensive model in Lehmer et al. (2020). The real variability is almost certainly larger than shown in Figure 8, as only narrow ranges of the key parameters were used.

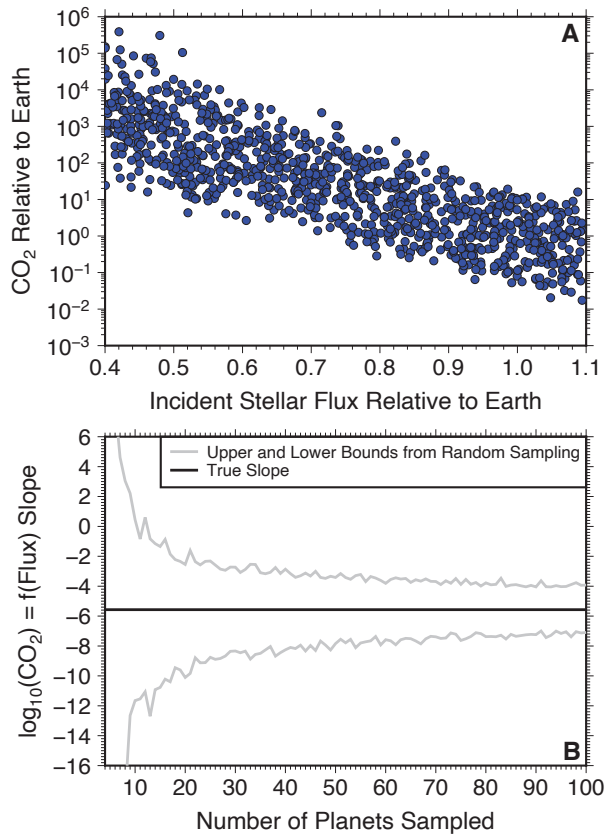

Figure 8: Atmospheric $CO_2$ as a function of incident stellar flux relative to what the modern-day Earth receives for 1000 modeled planets where key parameters for the carbonate-silicate cycle are randomly varied as described in the text (A). A given number of planets, ranging from 5 to 100, are then randomly drawn from the full set of 1000 models and a linear fit to their $\log_{10}(P)$ as a function of incident stellar flux is performed. The resulting range of slopes from this linear fitting of a random sample of planets is plotted as a function of the number of planets sampled, along with the "true" slope obtained from fitting the entire sample (B).

Turbet et al. (2019) proposed a different but related test of the carbonate-silicate cycle based on the onset of runaway greenhouse climates inward of the inner edge of the habitable zone (see also Schlecker et al. 2023). Planets inward of the habitable zone inner edge would experience runaway greenhouse climates with steam atmospheres. Such a climate state inflates the planetary radius measurably, meaning a shift from inflated to not inflated radii could pinpoint the inner edge of the habitable zone. At the same time, the carbonate-silicate cycle would predict a large jump in atmospheric $CO_2$ across the habitable zone inner edge, from low levels inside the habitable zone but near the inner edge, to Venus-like thick $CO_2$ atmospheres inward of the habitable zone inner edge (Turbet 2020). The approach of looking for major climate and atmosphere transitions across the habitable zone inner edge is probably more feasible than detecting an atmospheric $CO_2$ versus stellar flux trend. Lustig-Yaeger et al. (2022) showed that changes in transmission spectra with atmospheric $CO_2$ are small, requiring up to 100 transits per planet to resolve with JWST. Detecting a full population level trend of $CO_2$ versus flux is therefore not feasible, though null hypotheses instead could be tested. Moreover, atmospheric $CO_2$ partial pressure may be difficult even with HWO, as recent studies testing retrieval methodologies find Earth-like $CO_2$ levels difficult to constrain (Damiano & Hu 2022; Hall et al. 2023; Robinson & Salvador 2023). Looking for the transition to runaway greenhouse climates at the habitable zone inner edge is feasible with basic transit observations, rather than more intensive transmission spectroscopy (Schlecker et al. 2023). However, just observing runaway greenhouse climates at the habitable zone inner edge would also not test whether climate regulation throughout the habitable zone is a widespread process. Ultimately all of the approaches discussed here should be followed to test the idea of the habitable zone and climate regulation via the carbonate-silicate cycle, as they are complementary and constrain different aspects of the problem.

If we are able to develop a large dataset of characterized rocky planet atmospheres across the habitable zone in the coming decades, additional information about rocky planet geology could potentially be gleaned. One of these pieces of information is active volcanism and volatile cycling as a function of planet age. The carbonate-silicate cycle relies on planets having active volcanism and volatile cycling, either through plate tectonics (e.g. Sleep & Zahnle 2001) or stagnant-lid tectonics (e.g. Foley & Smye 2018; Valencia et al. 2018). However, once volcanism on a planet ceases due to mantle cooling, so does an active carbonate-silicate cycle. It is not clear how atmospheric $CO_2$ would evolve on a planet after the cessation of volcanism; Foley (2019) assumed the climate would cool as erosion and weathering would continue until topography is smoothed out and surface rocks are fully weathered. If this is true, then planets without active volcanism may have low atmospheric $CO_2$ levels. One could test this idea by looking at whether the expected trend of atmospheric $CO_2$ with stellar flux breaks down moving from a sample of younger planets to a sample of older planets, with the older planets showing lower atmospheric $CO_2$ and potentially a weaker relationship with flux. In fact, any breakdown of the expected trend for increasing planet age could be a sign of the cessation of the carbonate-silicate cycle feedback, regardless of whether older planets show lower than expected $CO_2$, higher than expected, or just increased random variability in either direction, given the uncertainty of how atmospheric $CO_2$ will respond to the end of volcanism.

In addition to planets geologically "dying" as they age, future work will likely identify additional factors that can cause planets to lose their carbonate-silicate cycle feedbacks. In a similar way to the discussion above, these factors can also be tested by looking to see if planets with these factors preferentially show larger deviations from the expected $CO_2$-flux trend. As such, a large dataset of rocky planets atmospheres covering as wide a range of environments and geological characteristics as possible will enable significant advances in our understanding of rocky planet evolution. The search for habitable exoplanets and signs of life therefore dovetails nicely with the goal of learning about rocky exoplanet geological evolution. However, even planets that are not likely to be habitable could still be valuable in helping constrain planetary evolution, so such planets should not be overlooked when prioritizing future observations.

# SUMMARY

The number of rocky exoplanets discovered to date dwarfs the number of rocky bodies in our solar system, thereby massively enhancing our sample size of rocky planet geological outcomes. However, in comparison to our study of solar system bodies, the information we can get from rocky exoplanets is severely limited, and typically only indirectly constraints the geological features or processes of these planets. It is therefore questionable how much we will actually be able to learn about rocky planet geological evolution from exoplanets. This chapter focused on a few aspects of rocky exoplanet geology that can potentially be constrained with current or near future observational capabilities. However, some of the most promising techniques involve looking at planets in exotic, compared to those in solar system, states or situations, or carefully extracting constraints from observations that are seemingly only loosely related geological processes. Planetary scientists looking to learn about the geological evolution of exoplanets will therefore need to shift perspective to the types of planets that are most readily observed, even if they are drastically different than Earth or other solar system planets.

One example is the most promising ways to directly constrain the composition of rocky exoplanets involve planets that have either already been destroyed, in the case of white dwarf pollution (see Veras et al. (20XX, this volume); Xu et al. (20XX, this volume)), or are being destroyed, in the case of disintegrating planets. Lava planets with magma oceans at the sub-stellar points or hot, but solid, airless planets are also promising avenues for direct characterization of the planet's crust or interior composition. Determining the range of exoplanet compositions is a vital goal in assessing all aspects of their geological evolution. Currently models of rocky exoplanet compositions are typically based on the abundances of refractory elements in their host stars, assuming that these compositions will be linked. While this is a reasonable approach, it is important to find ways directly measure rocky exoplanet compositions, to help benchmark these models and inform the diversity of rocky exoplanets.

Another example focused on how planets that lack atmospheres entirely can actually inform volcanic outgassing history, and potentially interior volatile inventory, of rocky planets. Airless rocky planets may seem like a disappointing discovery, in particular in the framework of astrobiology and searching for biosignatures. However, they are scientifically important for helping map the "cosmic shoreline" where planets are able to retain atmospheres, which is itself astrobiologically important. Moreover, as discussed in this chapter, an airless rocky planet constrains volatile outgassing rates, both at the present day and potentially over the planet's history, if atmospheric escape rates can be constrained. An airless planet must have escape rates that exceed outgassing rates at the present day, otherwise an atmosphere would be accumulating. Moreover, during the planet's history escape must have been able to remove whatever atmosphere may have formed during the planet's evolution. As a result, there is a clear need for collaboration between studies of atmospheric escape and stripping by stellar winds and studies of rocky planet interior evolution. Coupling these two processes in models of planetary evolution is essential for mapping which geological characteristics (like planet composition, starting volatile abundances, or interior oxidation state) can be ruled out by the finding of a planet lacking any atmosphere.

While current atmospheric characterization for rocky exoplanets is mostly only able to determine whether an atmosphere is present or not, future missions should allow more detailed characterization of atmospheric compositions. Such abilities will open up a wide range of new questions that can be answered about rocky planet evolution. This chapter focused on the possibility of testing whether the carbonate-silicate cycle climate feedback operates on rocky exoplanets in the same way it is thought to stabilize Earth's climate. Such a test will require measuring atmospheric $CO_2$ concentrations for 10s of rocky exoplanets, given the likely intrinsic variability on planets for weathering and outgassing. Moreover, if such a large dataset of rocky planet atmosphere characterizations can be built, additional geological features can potentially be inferred. For example, the typical lifetime of volcanism can potentially be constrained if it is found that beyond a certain range of ages the relationship between atmospheric $CO_2$ and incident stellar flux expected for an active carbonate-silicate cycle breaks down.

The constraints discussed in this chapter are basic. However, given how little is confidently known about rocky exoplanet geology, any additional constraints that can be developed still represent a significant advance. Whether we can go beyond the very basic constraints towards assessing things like the tectonic mode of a rocky exoplanet or directly observing active volcanism remains to be seen. Detecting volcanism, though, appears promising. Sulfate aerosols have been proposed as a potential marker of active volcanism (Kaltenegger et al. 2010). Explosive volcanism on Earth launches sulfate aerosols into the stratosphere, where they can exist for months to years before raining out. Stratospheric sulfate aerosols are readily detectable in transmission spectroscopy with instruments like JWST, and seeing a time-varying signature would plausibly indicate active volcanic eruptions on a planet (Misra et al. 2015). Ostberg et al. (2023) proposed an analogous method for detecting volcanism with direct imaging. The applicability of the sulfate aerosol volcanism signature, though, depends on the broader context of the planet, like the mantle and atmosphere composition, while detectability further depends on atmosphere composition and structure. In the end coupled models of atmospheric chemistry and outgassing would likely be needed to determine how well any proposed signature can be used as a marker of active volcanism. Nevertheless, any potential signature of exovolcanism is exciting and should be pursued, as it would be highly relevant for assessing exoplanet geology and habitability.

Inferring tectonic mode is likely to be even more difficult. There is no known unique signature of either plate tectonics or stagnant-lid tectonics that can be reasonably observed by proposed future telescopes. That the theory of plate tectonics was not developed until the 1960s despite hundreds of years of directly studying Earth's geology is a sobering reminder of the difficulty of determining tectonic mode. However, the more we can learn about rocky planets in any context, the more potential avenues for placing any type of constraint on tectonics. While there may be no one "smoking gun" signature, an accumulation of different lines of evidence all pointing towards one tectonic mode may eventually allow for strong inferences of exoplanet tectonics. The same logic holds for any other key aspect of exoplanet geology we might wish to learn about. The future will clearly require interdisciplinary, collaborative work to combine the different observations we are able to make into constraints on exoplanet geology.

# ACKNOWLEDGEMENTS

I thank Dimitri Veras and Josh Krissansen-Totton for thorough reviews that greatly improved the final paper. The Center for Exoplanets and Habitable Worlds is supported by the Pennsylvania State University and the Eberly College of Science.